\documentclass{aa}
\usepackage{natbib}
\bibpunct{(}{)}{;}{a}{}{,} 

\usepackage{lscape}

\usepackage{placeins}
\usepackage{graphicx}
\usepackage{amsmath}
\usepackage{txfonts}
\usepackage{color}
\usepackage[dvipsnames]{xcolor}
\usepackage{subcaption}

\graphicspath{{Figures2/}}

\newcommand{\mdisk}[0]{\ensuremath{M_{\rm disk}}}
\newcommand{\mgas}[0]{\ensuremath{M_{\rm gas}}}
\newcommand{\mdust}[0]{\ensuremath{M_{\rm dust}}}

\newcommand{\alp}[0]{\ensuremath{\alpha_{\rm visc}}}
\newcommand{\tvisc}[0]{\ensuremath{t_{\rm visc}}}
\newcommand{\rgas}[0]{\ensuremath{R_{\rm CO,\ 90\%}}}
\newcommand{\rc}[0]{\ensuremath{R_{\rm c}}}

\newcommand{\macc}[0]{\ensuremath{\dot{M}_{\rm acc}}}
\newcommand{\rinit}[0]{\ensuremath{R_{\rm init}}}

\newcommand{\mstar}[0]{\ensuremath{M_{*}}}
\newcommand{\msun}[0]{\ensuremath{\mathrm{M}_{\odot}}}
\newcommand{\cyo}[0]{C$^{18}$O}
\newcommand{\xco}[0]{$^{13}$CO}
\newcommand{\gdrat}[0]{\ensuremath{\Delta_{\rm gd}}}
\newcommand{\zetacr}[0]{\ensuremath{\zeta_{\rm cr}}}

\begin{document} 
    \title{CO isotopolog line fluxes of viscously evolving disks:} 
    \subtitle{Cold CO conversion insufficient to explain observed low fluxes}
    \titlerunning{CO isotopolog line fluxes of viscously evolving disks}
    
   \author{L. Trapman \inst{1,2}
          \and
          A.D. Bosman \inst{1,3}
          \and
          G. Rosotti \inst{1,4}
          \and
          M.R. Hogerheijde \inst{1,5}
          \and
          E.F. van Dishoeck \inst{1,6}
          }

   \institute{
            Leiden Observatory, Leiden University, Niels Bohrweg 2, NL-2333 CA Leiden, The Netherlands \\
            \email{ltrapman@wisc.edu}
        \and 
            Department of Astronomy, University of Wisconsin-Madison, 475 N Charter St, Madison, WI, 53706, USA
        \and
            Department of Astronomy, University of Michigan, 1085 S. University Ave, Ann Arbor, MI 48109, USA
        \and
            School of Physics and Astronomy, University of Leicester, Leicester LE1 7RH, UK            
        \and
            Anton Pannekoek Institute for Astronomy, University of Amsterdam, Science Park 904, 1090 GE Amsterdam, The Netherlands
        \and
            Max-Planck-institute f\"{u}r Extraterrestrische Physik, Giessenbachstra{\ss}e, D-85748 Garching, Germany
            }
   \date{Received xx; accepted yy}

  \abstract
   {Protoplanetary disks are thought to evolve viscously, where the disk mass - the reservoir available for planet formation - decreases over time as material is accreted onto the central star over a viscous timescale. 
   Observations have shown a correlation between disk mass and the stellar mass accretion rate, as expected from viscous theory. However, this happens only when using the dust mass as a proxy of the disk mass; the gas mass inferred from CO isotopolog line fluxes, which should be a more direct measurement, shows no correlation with the stellar mass accretion rate.}
   {We investigate how \xco\ and \cyo\ $J=3\,-\,2$ line fluxes, commonly used as gas mass tracers, change over time in a viscously evolving disk and use them together with gas disk sizes
   to provide diagnostics of viscous evolution. In addition, we aim to determine if the chemical conversion of CO through grain-surface chemistry combined with viscous evolution can explain the CO isotopolog observations of disks in Lupus.} 
   {We ran a series of thermochemical DALI models of viscously evolving disks, where the initial disk mass is derived from observed stellar mass accretion rates. }
    {While the disk mass, \mdisk, decreases over time, the \xco\ and \cyo\ $J=3\,-\,2$ line fluxes instead increase over time due to their optically thick emitting regions growing in size as the disk expands viscously. The \cyo\ 3-2 emission is optically thin throughout the disk for only for a subset of our models $(M_*\leq0.2\ \msun\ \mathrm{and\ } \alp\geq10^{-3}\ \mathrm{,\ corresponding\ to}\ \mdisk(t=1\ \mathrm{Myr})\leq10^{-3}\ \msun)$ . For these disks the integrated \cyo\ flux decreases with time, similar to the disk mass. 
    Observed \xco\ and \cyo\ 3-2 fluxes of the most massive disks $(\mdisk \gtrsim 5\times10^{-3}\ \msun)$ in Lupus can be reproduced to within a factor of $\sim2$ with viscously evolving disks in which CO is converted into other species through grain-surface chemistry with a moderate cosmic-ray ionization rate of $\zetacr\sim10^{-17}\ \mathrm{s}^{-1}$. 
    The \cyo\ 3-2 fluxes for the bulk of the disks in Lupus (with $\mdisk \lesssim 5\times10^{-3}\ \msun$) can be reproduced to within a factor of $\sim2$ by increasing $\zetacr\ \mathrm{to}\ \sim 5\times10^{-17} - 10^{-16}\  \mathrm{s}^{-1}$, although explaining the stacked upper limits requires a lower average abundance than our models can produce. In addition, increasing \zetacr\ cannot explain the observed \xco\ fluxes for lower mass disks, 
    which are more than an order of magnitude fainter than what is predicted.
    In our models the optically thick \xco\ emission originates from a layer higher up in the disk $(z/r\sim0.25-0.4)$ where photodissociation stops the conversion of CO into other species. Reconciling the \xco\ fluxes of viscously evolving disks with the observations requires either efficient vertical mixing or low mass disks ($\mdust \lesssim 3\times10^{-5}\ \msun$) being much thinner and/or smaller than their more massive counterparts.
    }
   {The \xco\ model flux predominantly traces the disk size, but the \cyo\ model flux traces the disk mass of our viscously evolving disk models if chemical conversion of CO is included. The discrepancy between the CO isotopolog line fluxes of viscously evolving disk models and the observations suggests that CO is efficiently vertically mixed or that low mass disks are smaller and/or colder than previously assumed.}
   
   \maketitle
%
\section{Introduction}
\label{sec: introduction}

With over 4200 exoplanets detected\footnote{\url{http://www.exoplanet.eu}} in a multitude of planetary systems, it has become clear that the formation of planets around young stars is a common occurrence (see, e.g., \citealt{Borucki2011,WinnFabrycky2015,Morton2016}). Understanding the processes behind the formation of planets has proven challenging. A key ingredient is the total amount of material available in the protoplanetary disks in which these planets form and grow (see, e.g., \citealt{Benz2014,Armitage2015,Mordasini2018}). 
The disk mass determines how much raw material can be accreted onto the forming planets. 
The disk mass is also not static as accretion onto the central star slowly decreases its mass over time. Combined, the disk mass and the stellar mass accretion rate determine the lifetime of the disk and therefore set an upper limit to the duration of planet formation.
Determining disk masses and how they evolve over time is therefore crucial for our understanding of planet formation.

It is commonly assumed that protoplanetary disks evolve viscously (see, e.g., \citealt{LyndenBellPringle1974,ShakuraSunyaev1973}). Viscous stresses in the disk transport angular momentum outward, causing the outer parts of the disk to spread out. To conserve angular momentum, this causes material to be transported inward, where it is accreted onto the central star. While still debated, the physical process behind this effective viscosity is commonly assumed to be the magnetorotational instability (see, e.g.,\citealt{BalbusHawley1991,BalbusHawley1998}).
In the framework of viscous evolution, the disk mass, \mdisk, and the stellar mass accretion rate, \macc, follow a linear relation, with the disk mass being accreted onto the star on a viscous timescale $(\mdisk \sim \macc\tvisc)$.

Alternatively, it has been suggested that the angular momentum in disks can be removed by a magnetic disk wind rather than transported outward by viscous stresses (see, e.g \citealt{Turner2014}). This magnetic disk wind forms in the presence of a vertical magnetic field in the disk, and it is able to remove material from the disk surface, thus reducing the total angular momentum in the disk. What fraction of the angular momentum can be carried away by the disk wind is still a matter of debate (see, e.g., \citealt{Ferreira2006,Bethune2017,ZhuStone2017}). 
Magnetic disk winds have been detected in observations but predominantly in the inner part of the disk, and it remains unclear how much disk winds affect disk evolution (see, e.g., \citealt{Pontoppidan2011,Bjerkeli2016,Tabone2017, deValon2020}). 

Recently, several combined observing campaigns have performed large surveys of the full disk population, allowing the simultaneous study of properties of protoplanetary disks and the young stellar objects that host them. 
Several star-forming regions have  been covered by Atacama Large Millimeter/submillimeter Array (ALMA) disk surveys, providing high angular resolution observations of disk continuum emission and carbon monoxide (CO) rotational line emission (e.g., \citealt{ansdell2016,ansdell2017,ansdell2018,Barenfeld2016,barenfeld2017,Pascucci2016,Long2017,Cazzoletti2019,Cox2017,Cieza2019,Williams2019}).
Using spectra from the X-shooter spectrograph \citep{Vernet2011} at the ESO Very Large Telescope, stellar properties such as the stellar mass accretion rate have also been measured for a large fraction of the disk-hosting stars in star-forming regions such as Lupus, Chamaeleon I, and Upper Sco \citep{alcala2014,alcala2017,manara2017,Manara2020}.

Combining observations for the disk population in Lupus, \cite{Manara2016} found a correlation between the observed stellar mass accretion rate, \macc, and the disk dust mass, \mdust, derived from millimeter continuum emission. If an interstellar medium (ISM) gas-to-dust mass ratio $\gdrat = 100$ is assumed, the observations (\macc\ and $\mdisk = 100\times\mdust$) are consistent with viscous disks having evolved for 1-3 Myr, which matches the approximate age of the sources (see also \citealt{Rosotti2017}).
Interestingly, they find no correlation between \macc\ and the disk gas mass, \mgas, derived from \xco\ $J=3\,-\,2$ and \cyo\ $J=3\,-\,2$ line fluxes. This seems to contradict their first finding, suggesting that the disk gas mass, which is expected to make up most of the total disk mass, is not related to the stellar mass accretion rate. 

The cause for this discrepancy might lie with the tracer used to measure the disk gas mass. For most disks the gas masses derived from optically thin CO isotopologs such as \xco\ and \cyo\ are found to be low compared to their dust mass, with $\gdrat = \mgas/\mdust \approx 1-10$ for most disks (see, e.g.,  \citealt{ansdell2016,miotello2017,Long2017}). \textit{Herschel} Space observatory observations of the hydrogen deuteride (HD) $J=1\,-\,0$ rotational line toward a handful of disks have provided an independent measurement of the disk gas mass \citep{Bergin2013,McClure2016,Trapman2017,Kama2020}. These observations find a gas-to-dust mass ratio of $\sim100$, suggesting that the low CO-based gas masses are a sign that disks are underabundant in CO. This underabundance is in addition to well-understood processes such as CO freeze-out and photodissociation. Several processes have been suggested to explain the extra underabundance of CO, such as chemical conversion of CO in the gas or on the grains into more complex species (e.g., \citealt{Aikawa1997,Bergin2014,Furuya2014,Yu2016,Yu2017,DodsonRobinson2018,Bosman2018b,Schwarz2018}) or the locking of CO in larger bodies (see, e.g., \citealt{Bergin2010,Bergin2016,Kama2016,Krijt2018}). 

In this work we use an alternative approach to investigate the lack of a correlation between \macc\ and the CO-based \mgas, by taking a step back and examine what \xco\ and \cyo\ line fluxes are expected for viscously evolving disks. Over time the disk spreads out, increasing the emitting region,
and the disk mass decreases, lowering the \xco- and \cyo-column densities, both of which affect the resulting line fluxes. 
Furthermore, by using the initial gas masses that can explain the observed stellar mass accretion rates, we can examine if the observed \xco\ and \cyo\ line fluxes are consistent with viscous evolution.

Recently, \cite{Trapman2020} used the same modeling framework to examine if observed gas outer radii of disks in the Lupus and Upper Sco star-forming regions can be explained with viscous evolution (see also \citealt{ansdell2018,barenfeld2017}). They showed that gas outer radii of disks in Lupus are consistent with viscously evolving disks that start out small, meaning an initial characteristic radius of $\sim10$ AU, and that have a low viscosity $(\alp = 10^{-4} - 10^{-3})$. 
Combining their results with our analysis of the CO isotopolog line fluxes, we can examine if disks are in agreement with viscous theory in terms of both their size and their mass.

The structure of this work is as follows:
In Section \ref{sec: Model setup} we discuss the setup and initial conditions of our models. Here we also outline how we implement the chemical conversion of CO through grain-surface chemistry.
In Section \ref{sec: results} we first show how \xco\ and \cyo\ intensity profiles and integrated line fluxes change over time in a viscously evolving disk and how they shift if grain-surface chemistry converts CO into other species. Next we compare our models to observations in Lupus and discuss the cosmic-ray ionization rates that are required to match the observed fluxes.
In Section \ref{sec: discussion} we look in more detail at the \xco\ and \cyo\ fluxes that are overproduced by our models and we discuss the impact of vertical mixing and alternative explanations such as small disks.
We conclude in Section \ref{sec: conclusions} that reconciling the \xco\ and \cyo\ fluxes of viscously evolving disks models with the observations requires either that CO is efficiently mixed vertically or that low mass disks are small.

\section{Model setup}
\label{sec: Model setup}

For setting up the initial conditions and time evolution of our models, we follow the same approach as presented \cite{Trapman2020}. For completeness we reiterate the steps in this approach here.

Based on observed stellar accretion rates we calculate what the initial disk mass must have been, assuming that the disk has evolved viscously. From the initial disk masses we compute the surface density profile analytically at 10 consecutive disk ages. At each time step the thermochemical code  \underline{D}ust \underline{a}nd \underline{Li}nes (\texttt{DALI}; \citealt{Bruderer2012,Bruderer2013}) is used to calculate the abundances, excitation and temperature of the disk and the model is ray-traced to obtain \xco\ $J=3\,-\,2$ and \cyo\ $J=3\,-\,2$ line fluxes.
This modeling approach was previously used by \cite{Trapman2020} to study the evolution of measured gas disk sizes.

\subsection{Viscous evolution of the surface density}
\label{sec: viscous surface density}

While the physical processes underlying the viscosity in disks are still an open question, 
it is common to describe the kinetic viscosity $\nu$ using the dimensionless parameter \alp, defined as $\nu = \alpha c_s H$, where $c_s$ is the sound speed and $H$ is the height above the midplane (the $\alpha-$disk formalism, see \citealt{ShakuraSunyaev1973,Pringle1981}). In this formalism a self-similar solution for the surface density $\Sigma$ can be calculated \citep{LyndenBellPringle1974,Hartmann1998}

\begin{equation}
\label{eq: surface density}
    \Sigma_{\rm gas}(R) = \frac{\left(2-\gamma\right)\mdisk(t)}{2\pi\rc(t)^2} \left( \frac{R}{\rc(t)} \right)^{-\gamma} \exp\left[ -\left(\frac{R}{\rc(t)}\right)^{2-\gamma}\right],
\end{equation}
where $\gamma$ enters by assuming that the viscosity varies radially as $\nu\propto R^{\gamma}$ and \mdisk\ and \rc\ are the disk mass and the characteristic radius, respectively. 

\begin{figure}
    \centering
    \includegraphics[width=\columnwidth]{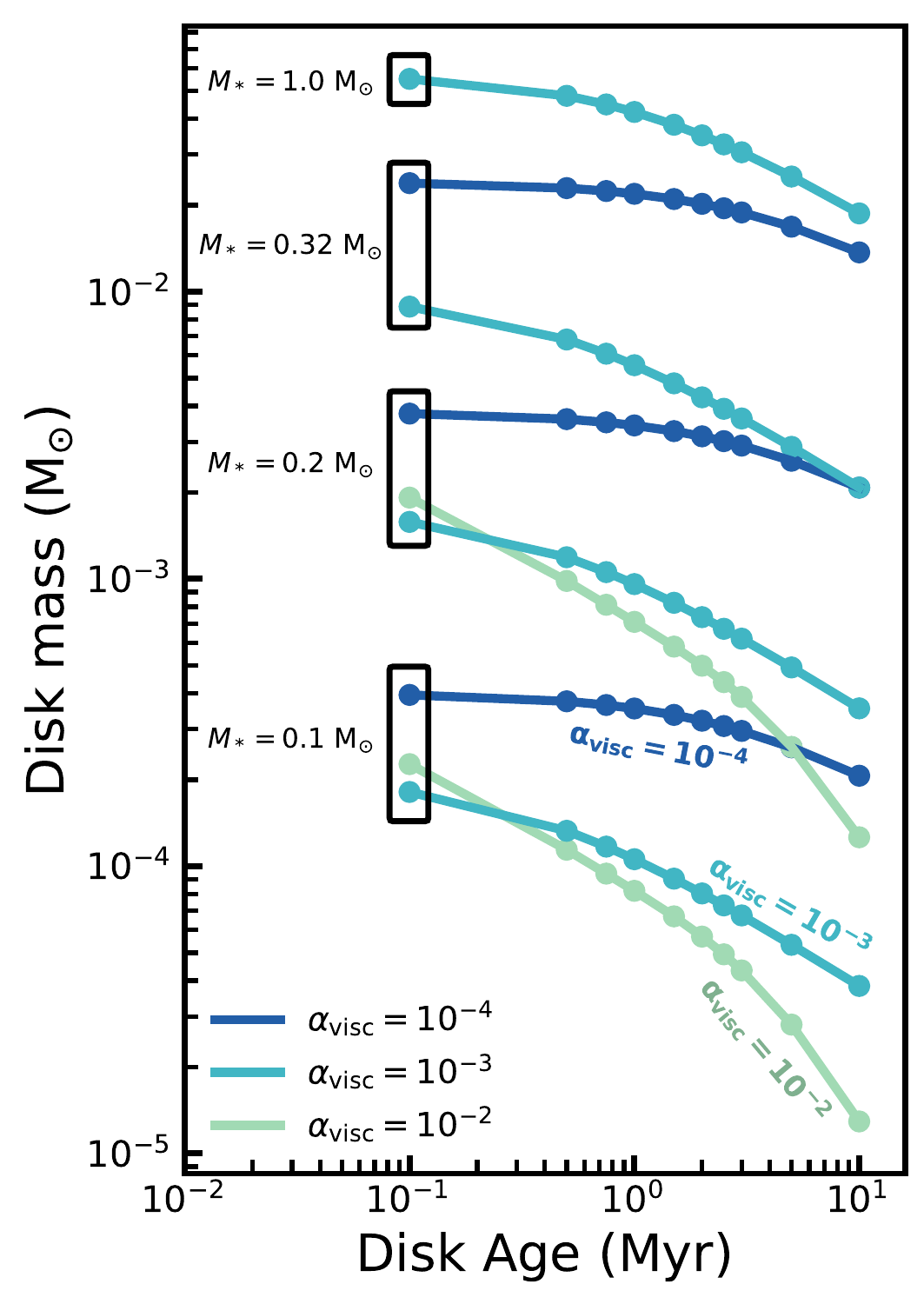}
    \caption{\label{fig: disk mass evolution} Time evolution of the disk mass, \mdisk, of our models. Colors indicate the viscous alpha, \alp, of the models. The black boxes show which stellar mass, and therefore accretion rate, was used to calculate the initial disk mass of the models (see Table \ref{tab: model initial conditions}). We note that we exclude models with $\alp = 10^{-2}$ for $\mstar = [0.32,1.0]\ \msun$ as the gas disk sizes of these models can be ruled out by observations of disks in Lupus \citep{Trapman2020}. We also note that models with $\mstar=1.0\ \msun$ and $\alp  = 10^{-4}$ are much more massive $(\mgas \gtrsim2\times10^{-1}\ \msun)$ than the disks observed in Lupus (cf. \citealt{ansdell2016}; see Figure A.1 in \citealt{Trapman2020}).}
\end{figure}

The evolution of the surface density depends on how \mdisk\ and \rc\ change over time (see, e.g., \citealt{Hartmann1998}):
\begin{align}
\label{eq: mass time evolution}
\mdisk(t) &= \mdisk(t=0) \left(1 + \frac{t}{\tvisc} \right)^{-\frac{1}{[2(2-\gamma)]}}\\
          &= \mdisk(t=0) \left(1 + \frac{t}{\tvisc} \right)^{-\frac{1}{2}}\\
\rc(t)    &= \rc(t=0) \left(1 + \frac{t}{\tvisc} \right)^{\frac{1}{(2-\gamma)}}\\
          &= \rc(t=0) \left(1 + \frac{t}{\tvisc} \right).
\end{align} 
Here \tvisc\ is the viscous timescale. For the second step and in the rest of this work we have assumed $\gamma = 1$. For a typical temperature profile this corresponds to the case 
where \alp\ stays constant with radius. 
The time evolution of \mdisk\ for our models can be seen in Figure \ref{fig: disk mass evolution}.

\subsection{Initial disk mass and disk size}
\label{sec: initial condition}

Over time the mass of a disk decreases as material is accreted onto the central star. For a disk that is viscously evolving with a constant \alp\ the relation between the initial disk mass $\mdisk(t=0)$ and the stellar mass accretion rate, \macc, at a given time $t$ can be written such that the initial disk mass is a function of \macc(t) (see, e.g., \citealt{Hartmann1998})

\begin{equation}
\label{eq: init disk - accretion}
    \mdisk(t=0)\ \ = 2\ \tvisc\ \macc(t)\ \left(\frac{t}{\tvisc} + 1 \right)^{3/2}.
\end{equation}

Stellar mass accretion rates have now been measured for multiple star-forming regions such as Lupus, Chamaeleon I, and Upper Sco \citep{alcala2014,alcala2017,manara2017,Manara2020}. Based on these accretion rates we can calculate what the initial disk masses would have been, given the viscous timescale and the average age of the star-forming region.

To determine the initial disk masses of the models, our approach is as follows. We take four stellar mass accretion rates ($4\times10^{-12}, 4\times10^{-11},2\times10^{-9}\ \mathrm{and}\ 10^{-8}\ \msun\ \mathrm{yr}^{-1}$) that span the range of the observations in Lupus (see, e.g., \citealt{alcala2014,alcala2017}). As observations have shown that \macc\ is correlated to \mstar, our selected \macc\ represent the average accretion rates for stars with $\mstar = 0.1, 0.2, 0.32\ \mathrm{,\ and}\ 1.0\ \msun$. Compared to \cite{Trapman2020} this includes one more set of models $(\mstar = 0.2\ \msun)$. For these stellar mass accretion rates we calculate what the initial disk mass must have been for three different viscous timescales, assuming an average age of 1 Myr for Lupus. The viscous timescales are calculated for three values of the dimensionless viscosity $\alp = 10^{-2}, 10^{-3}, 10^{-4}$, assuming a characteristic radius of 10 AU (a choice that is discussed below) and a disk temperature $T_{\rm disk}$ of 20 K (see, e.g., equation 37 in \citealt{Hartmann1998}) 

\begin{equation}
\frac{\tvisc}{\mathrm{yr}} = \frac{\rc^2}{\nu} \simeq 8\times10^{4} \left(\, \frac{10^{-2}}{\alp}\,\right) \left(\,\frac{\rc}{10\ \mathrm{AU}}\,\right)\left(\,\frac{M_*}{0.5 \ \mathrm{M}_{\odot}} \,\right)^{1/2} \left(\,\frac{10\ \mathrm{K}}{T_{\rm disk}}\,\right).
\end{equation}

The resulting initial disk masses, 12 in total, are summarized in Table \ref{tab: model initial conditions}. Figure \ref{fig: disk mass evolution} shows how these disk masses decrease over time. In our analysis we exclude three models. \cite{Trapman2020} showed that the gas disk sizes measured from models with $\alpha = 10^{-2}$ and $\mstar = [\,0.32,1.0\,]\ \msun$ can be ruled out based on observed gas disk sizes in Lupus. 
We also exclude the model with $\mstar = 1.0\ \msun\ \mathrm{and}\ \alpha = 10^{-4}$, which has an initial disk mass of $\mdisk = 2.6\times10^{-1}\ \msun$ . At all points during its evolution, this model is factor of five more massive than the most massive disk observed in Lupus, but consistent with Class 0/1 disk masses (cf. Figure A.1 in \citealt{Trapman2020}; see also \citealt{Tychoniec2018,Tychoniec2020,Tobin2020}).

For the initial disk size we assume that disks start out small, with an initial characteristic disk size $\rc(t=0) = 10$ AU. Recently, \cite{Trapman2020} showed that the observed gas outer radii of Class II disks in Lupus can be explained by viscously evolving disks that start out small ($\rc(t=0) = 10\ \mathrm{AU}$) and that have a low viscosity $(\alp = 10^{-4} - 10^{-3})$. More importantly, they show that the bulk of the observed gas outer radii cannot be explained by disks that start out large ($\rc(t=0) \geq 30\ \mathrm{AU}$).

Observational constraints on the disk size of young Class 1 and 0 objects have only recently become available. In their VANDAM II survey, \citealt{Tobin2020} presented ALMA 0.87 millimeter observations of 330 protostars in Orion at a resolution of $\sim0\farcs1$ ($\sim40$ AU in diameter). Their observations suggest that the majority of disks are initially small ($\sim37-45\ \mathrm{AU}$ in radius), at least as seen in the emission of the millimeter dust. It is worth mentioning that their radii are defined as half of the full with at half maximum (FWHM) of a 2D Gaussian fit to the observations, which is not the same as the characteristic radius of the disk. 
For the gas there is similar evidence that disks start out small, albeit from a smaller sample. \cite{Maret2020} presented NOEMA observations of 16 Class 0 protostars as part of the CALYPSO large program. They found only two sources that show a Keplerian disk larger than $\sim 50$ AU. This suggests that stars with a large Keplerian disk at a young age, such as those found for VLA 1623 \citep{Murillo2013}, are uncommon. 
We therefore adopted an initial disk size of $\rinit = 10\ \mathrm{AU}$ for our models.

\begin{table}[htb]
    \centering
    \def\arraystretch{1.4}
    \caption{\label{tab: model initial conditions} Initial conditions of our \texttt{DALI} models}
    \begin{tabular}{lc|ccc}
    \hline
    \hline
     \mstar & \macc & \multicolumn{3}{c}{\alp}\\
     $(\mathrm{M}_{\odot})$ & $(\mathrm{M}_{\odot}\ \mathrm{yr}^{-1})$ & $10^{-2}$ & $10^{-3}$ & $10^{-4}$\\
    \hline
     & & \multicolumn{3}{c}{$\mdisk\ (\mathrm{M}_{\odot})$ at $t=0$ }\\
    \hline
      0.1  & $4\times10^{-11}$ & $4.5\times10^{-4}$ & $2.1\times10^{-4}$ & $4.1\times10^{-4}$ \\
      0.2  & $4\times10^{-10}$ & $3.6\times10^{-3}$ & $1.8\times10^{-3}$ & $3.9\times10^{-3}$ \\
      0.32 & $2\times10^{-9}$  & {\small $\mathit{2.0\times10^{-2}}$} & $1.0\times10^{-2}$ & $2.5\times10^{-2}$ \\
      1.0  & $1\times10^{-8}$  & {\small$\mathit{6.9\times10^{-2}}$} & $5.9\times10^{-2}$ & {\small $\mathit{2.6\times10^{-1}}$} \\
    \hline
    \end{tabular}
    \captionsetup{width=.90\columnwidth}
    \caption*{\footnotesize{\textbf{Notes.} 
    Disk masses in italic are not included in our analysis (see Section \ref{sec: initial condition}).
    The viscous timescale in the models varies approximately as $\tvisc \simeq 0.5\times\left(10^{-3}/\alp\right)\times10^{6}\ \mathrm{yr}$.}}
\end{table}

\subsection{The {\normalfont \texttt{DALI}} models}
\label{sec: DALI models}

Using the initial conditions discussed previously we computed how \mdisk\ decreases and \rc\ increases over time. For ten points in the lifetime of the disk, starting at 0.1 Myr and ending at 10 Myr, we computed the current surface density $\Sigma_{\rm gas}(t)$ of the disk model.

We used the thermochemical code \texttt{DALI} to compute CO isotopolog abundances and line fluxes of our disk models. 
\texttt{DALI} is a 2D physical-chemical code that computes the thermal and chemical structure for a given physical disk structure. 
For each $\Sigma_{\rm gas}(t)$ a Monte Carlo radiative transfer calculation is used determine the dust temperature and the internal radiation field from UV- to mm-wavelengths.
Next, the code computes atomic and molecular abundances by solving the time-dependent chemistry. The atomic and molecular excitation is then obtained using a non-LTE calculation. The heating and cooling balance is solved to calculate the gas temperature. These three steps were performed iteratively until a self-consistent solution is obtained. Raytracing of the model then yields line fluxes.
For a more detailed description of the code we refer the reader to Appendix A of \cite{Bruderer2012}.

To obtain a 2D (i.e., $R,z$) density structure for our models, we assumed that in the vertical direction the disk follows a Gaussian density distribution, as motivated by hydrostatic equilibrium (see, e.g., \citealt{ChiangGoldreich1997}). The height of the disk was assumed to follow a powerlaw of the form $H = R h = R h_c \left(R/R_{\rm c}\right)^{\psi}$, where $h_c$ is the opening angle at $R_{\rm c}$ and $\psi$ is the flaring angle.

As grains grow larger they decouple from the gas and settle toward the midplane of the disk (see, e.g., \citealt{dubrulle1995,DullemondDominik2004b,DullemondDominik2005,MuldersDominik2012,RiolsLesur2018}). The settling of large grains is included in our models in a simplified way by splitting the dust grains into two populations, following the approach in \cite{Andrews2011}. 
Most of the dust mass (90\%) is in large grains with sizes of 1-$10^3\ \mu$m. These large grains follow the gas radially, but vertically they are confined to the midplane as their scale height is reduced by a factor $\chi = 0.2$ with respect to the gas. A population of small grains (0.005-1 $\mu$m) make up the remaining 10\% of the dust mass in the disk. These small grains follow the distribution of the gas both radially and vertically. 

We assume that our disks encircle a typical T-Tauri star, which we represent with a blackbody with effective temperature for $T_{\rm eff} = 4000$ K. A second blackbody with $T_{\rm eff}  = 10000$ K is added to simulate the excess UV radiation released by material accreted onto the star. To compute the luminosity of this second component we follow \cite{kama2015} and assume that the gravitational potential energy of the accreted material is converted into radiation with a 100\% efficiency. 
These parameters are summarized in Table \ref{tab: model fixed parameters}.
We note that stellar evolution is not included in our models. Including stellar evolution can affect CO chemistry and its chemical conversion. Especially the higher stellar luminosity early on will increase the disk temperature and reduce the amount of CO freeze out (see e.g., the observations presented in \citealt{vtHoff2020a,vtHoff2020b}). 
A detailed study of the effect of stellar evolution on CO chemistry is available in \cite{Yu2016}.

\begin{table}[htb]
  \centering   
  \caption{\label{tab: model fixed parameters}\texttt{DALI} parameters of the physical model.}
  \begin{tabular*}{0.8\columnwidth}{ll}
    \hline\hline
    Parameter & Range\\
    \hline
     \textit{Chemistry}&\\
     Chemical age & 0.1-10 Myr$^{*,\dagger}$\\
     Volatile {[C]/[H]} & $1.35\times10^{-4}$\\
     Volatile {[O]/[H]} & $2.88\times10^{-4}$\\
     \textit{Physical structure} &\\ 
     $\gamma$ &  1.0\\ 
     $\psi$ & 0.15\\ 
     $h_c$ &  0.1 \\ 
     \rc & $10-3\times10^{3}$ AU$^{\dagger}$\\
     $M_{\mathrm{gas}}$ & $10^{-5} - 10^{-1}$ M$_{\odot}^{\dagger}$ \\
     Gas-to-dust ratio & 100 \\
     \textit{Dust properties} & \\
     $f_{\mathrm{large}}$ & 0.9 \\
     $\chi$ & 0.2 \\
     composition & standard ISM$^{1}$\\
     $f_{\rm PAH}$ & 0.001 \\
     \textit{Stellar spectrum} & \\
     $T_{\rm eff}$ & 4000 K + Accretion UV \\
     $L_{*}$ & 0.28 L$_{\odot}$  \\
     $L_{\rm X}$ & $10^{30}\ \mathrm{erg\ s^{-1}}$\\
     $T_X$ & $3.2\times10^{6}$ K \\
     $\zeta_{\rm cr}$ & $10^{-17}\ \mathrm{s}^{-1}$\\
     \textit{Observational geometry}&\\
     $i$ & 0$^{\circ}$ \\
     PA & 0$^{\circ}$ \\
     $d$ & 150 pc\\
    \hline
  \end{tabular*}
  \captionsetup{width=.75\columnwidth}
  \caption*{\footnotesize{\textbf{Notes.} $^*$The age of the disk is taken into account when running the time-dependent chemistry. $^{\dagger}$These parameters evolve with time. 
  $^{1}$\citealt{WeingartnerDraine2001}, see also Section 2.5 in \citealt{Facchini2017}. }}
\end{table}

\begin{figure}
    \centering
    \includegraphics[width=\columnwidth]{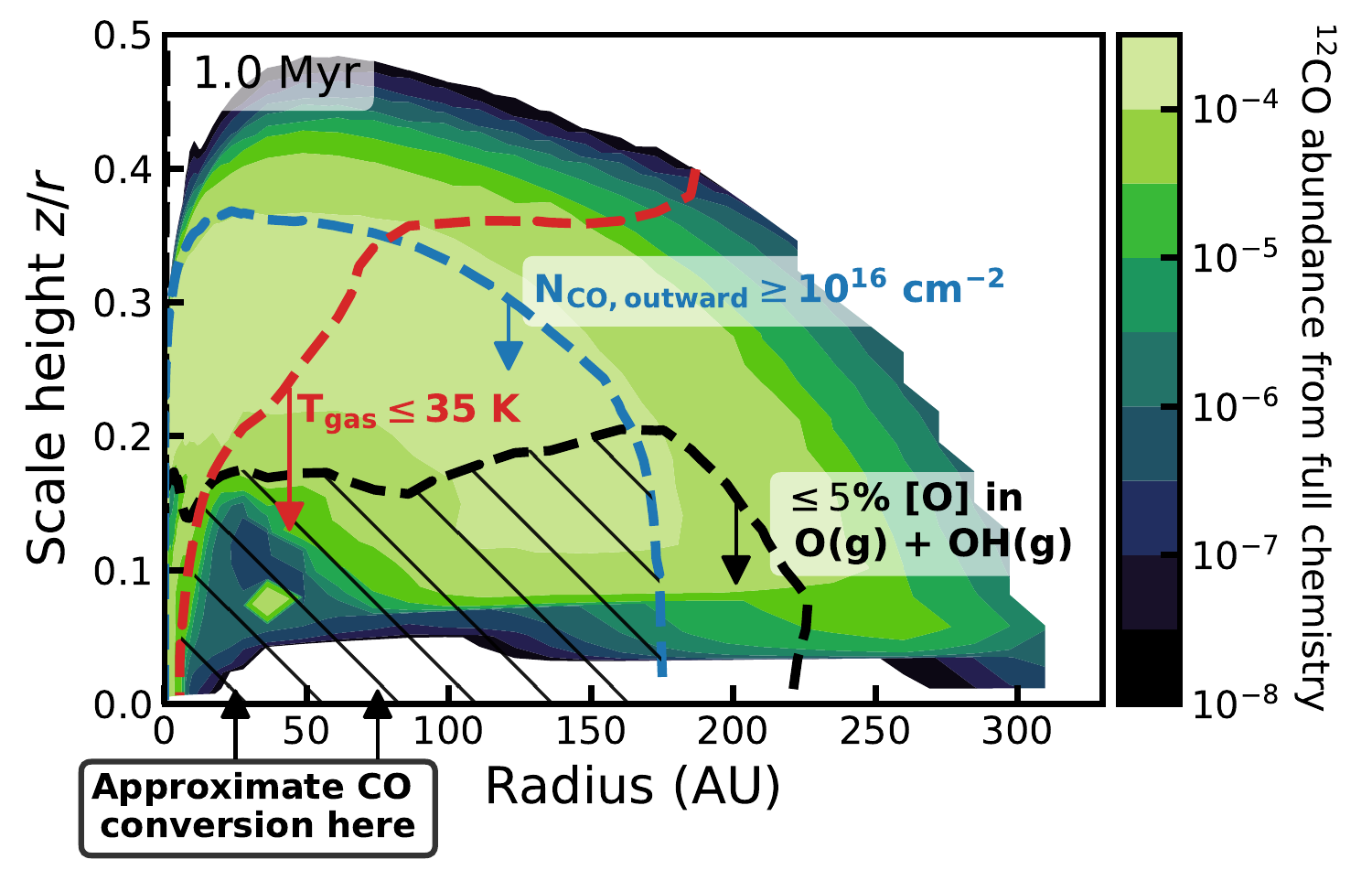}
    \caption{\label{fig: boundary approximation} CO abundance computed using the full chemical network in Bosman et al. (\citeyear{Bosman2018b}; see also \citealt{Cridland2020}) for the model with $\mstar = 0.32\ \msun\ \mathrm{and}\ \alp=10^{-3}$ at 1 Myr. The chemistry was calculated using a $\zetacr = 10^{-16}\ \mathrm{s}^{-1}$ to show where CO can be transformed. Taken from the same model, but with the chemistry calculated within \texttt{DALI}, the dashed blue line shows where the CO outward column is $10^{16}\ \mathrm{cm}^{-2}$. The black contour shows where 5\% of the total oxygen abundance is in OH(g) + O(g). The red contour shows where $T_{\rm gas} = 35\ \mathrm{K}$. The hatched region below the three dashed lines shows where the CO abundance is recalculated using the approximate grain-surface chemistry. }
\end{figure}

\begin{figure*}[!thb]
    \centering
    \begin{subfigure}{0.98\textwidth}
    \includegraphics[width=\textwidth]{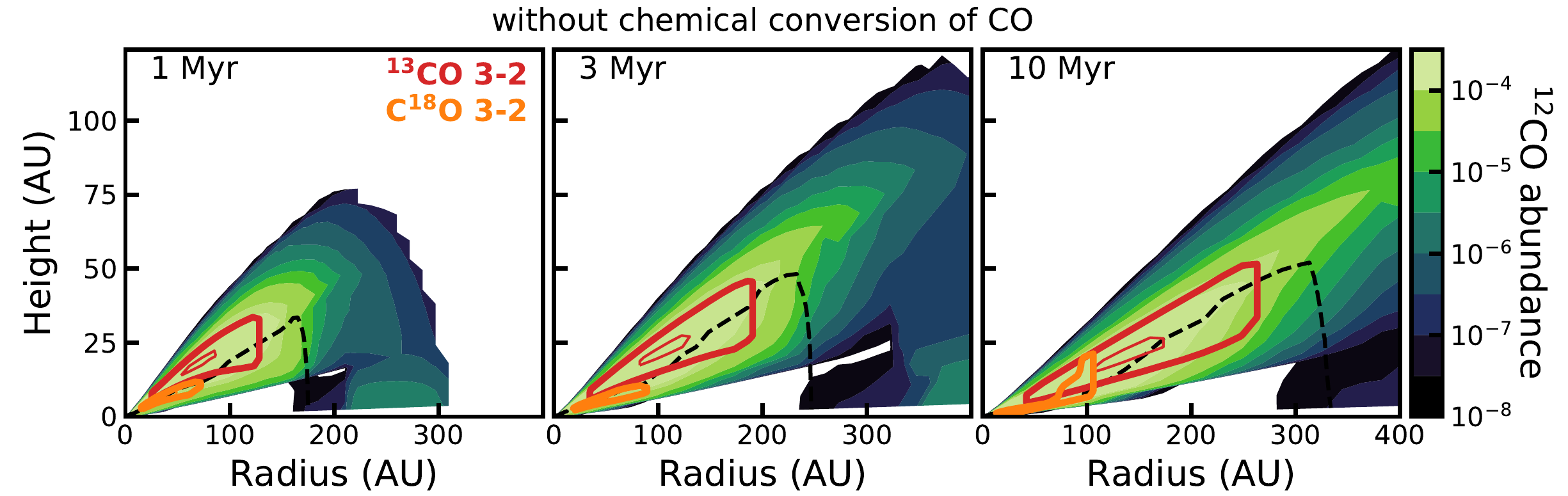}
    \end{subfigure}
    \begin{subfigure}{0.98\textwidth}
    \includegraphics[width=\textwidth]{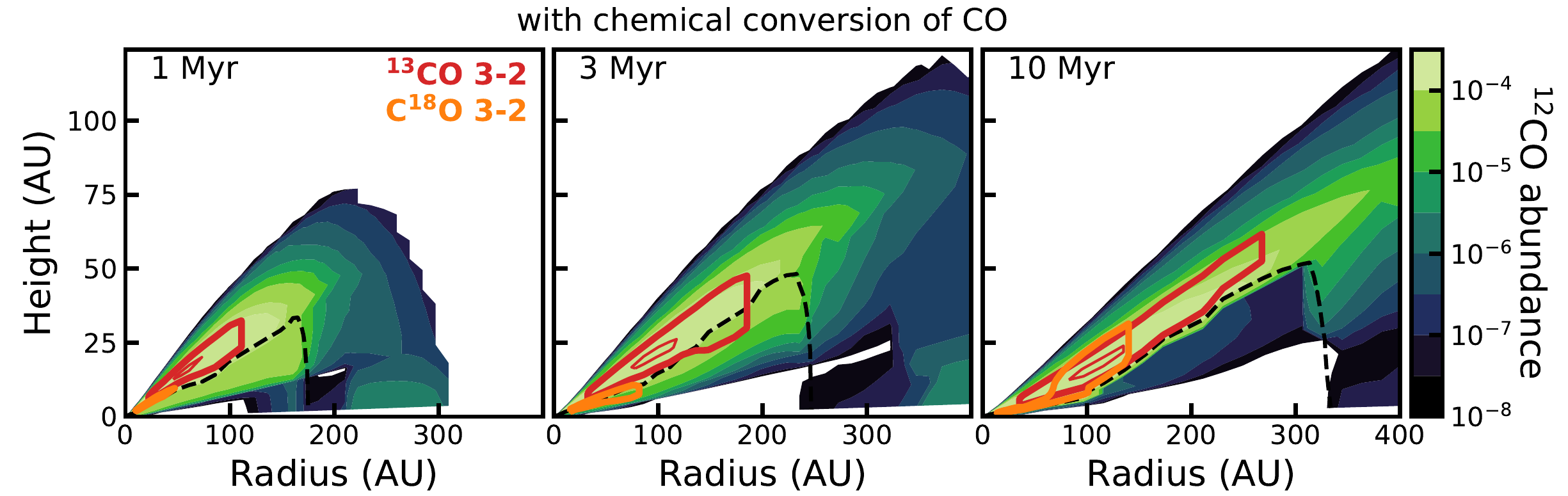}
    \end{subfigure}
    \caption{\label{fig: CO depletion in abundance} 
    Effect of the chemical conversion of CO on the CO abundance after 1, 3 and 10 Myr. The example model shown here has $M_* = 0.32\ \mathrm{M}_{\odot}$ and $\alp = 10^{-3}$. In the top panels CO abundances have been calculated using the DALI chemical network (see Section \ref{sec: chemical network}). In the bottom panels we have included the chemical conversion of CO into other species, as described in Section \ref{sec: grain surface chemistry}. 
    Colors show the CO abundance with respect to H$_2$, where white indicates CO/H$_2 \leq 10^{-8}$. 
    The dashed black line shows where $\leq5$\% of the total amount of oxygen is in O(g)+OH(g), the gas temperature $T_{\rm gas} \leq 35\ \mathrm{K}$ and outward CO column is $\geq10^{16}$ cm$^{-2}$.
    Grain-surface chemistry is calculated below this black dashed line.
    Red contours show the $^{13}$CO $J=3\,-\,2$ emitting region, enclosing 25\% and 75\% of the total $^{13}$CO flux. Similarly, the orange contours show the C$^{18}$O $J=3\,-\,2$ emitting region. }
\end{figure*}

\subsubsection{Isotopolog chemical network}
\label{sec: chemical network}

For less abundant isotopologs of CO such as $^{13}$CO, C$^{18}$O, and C$^{17}$O, isotope-selective photodissociation can have significant effect on their abundances (see, e.g., \citealt{Visser2009,Miotello2014}). Using the results from \citealt{Visser2009}, \cite{Miotello2014} expanded the standard chemical network of DALI, based on the UMIST 06 network \citep{woodall2007,Bruderer2012,Bruderer2013}, to include the isotope-selective photodissociation and chemistry of $^{13}$CO, C$^{18}$O, and C$^{17}$O. The chemical network includes the formation of H$_2$ on grains, freeze-out, thermal and nonthermal desorption of species, hydrogenation of simple species on ices, gas-phase reactions, photodissociation, X-ray- and cosmic-ray-induced reactions, polycyclic aromatic hydrocarbon (PAH) grain charge exchange and/or hydrogenation, and reactions with vibrationally excited H$_2^*$. A detailed description of the implementation of these reactions can be found in Appendix A.3.1 of \cite{Bruderer2012}. A full description of the isotopolog chemical network can be found in \cite{Miotello2014}.
For the initial volatile carbon and oxygen abundances, typical ISM values of [C]/[H]$_{\rm ISM} = 1.35\cdot10^{-4}$, [O]/[H]$_{\rm ISM} = 2.88\cdot10^{-4}$ were used \citep{Cardelli1996,Jonkheid2007,Woitke2009,Bruderer2012}. We take the isotope ratios to be [$^{12}$C]/[$^{13}$C] = 77 and [$^{16}$O]/[$^{18}$O] = 560 \citep{WilsonRood1994}.

Relevant for this work is that in the \texttt{DALI} chemical network CO is not further processed once it is frozen out on the grains. The only reaction included in the network that affects CO ice is desorption back into the gas phase. 

\subsubsection{CO conversion through grain-surface chemistry}
\label{sec: grain surface chemistry}

Recent observations have shown that a large fraction of protoplanetary disks have unexpectedly low \xco\ and \cyo\ line fluxes (see, e.g., \citealt{ansdell2016,miotello2017,Long2017}). Gas masses derived from these line fluxes using models that include freeze-out and photodissociation suggest that the bulk of the protoplanetary disks are gas poor, with gas-to-dust mass ratios $(\gdrat)$ on the order of $\gdrat \approx1-10$. For a few disks the gas mass has been determined independently using HD (see, e.g., \citealt{Bergin2013,McClure2016,Trapman2017,Kama2020}). These observations suggest instead that CO is underabundant in disks (see, e.g., \citealt{Favre2013,Du2015,Bergin2016,Kama2016,Zhang2019}). This requires some process currently not accounted for that removes CO from the gas phase. Two such processes have been suggested: The first proposes that when CO freezes out on grain it can become locked up in larger bodies, preventing it from reentering the gas phase (see, e.g., \citealt{Bergin2010,Bergin2016,Du2015,Kama2016,Krijt2018,Krijt2020}).
Radial drift of these larger bodies can move the frozen out CO to smaller radii (see, e.g \citealt{Booth2017}).
Several authors have also shown that grain-surface chemistry is capable of lowering the CO abundance in disks. In the gas collisions with He$^+$ can break apart CO molecules, allowing the available carbon to be locked up in hydrocarbons such as CH$_4$ and C$_2$H$_6$ (see, e.g., \citealt{Aikawa1997,Bergin2014,Schwarz2018}).
In the ice CO can be converted into other species such as CO$_2$ and CH$_3$OH on a timescale of several megayears (see, e.g., \citealt{Bergin2014,Yu2016,Yu2017,Bosman2018b,Schwarz2019}). 

Calculating the ``chemical conversion of CO'' requires a large chemical network. As expanding the network significantly increases the computation time of a thermochemical model, it is unfeasible to calculate the chemical conversion of CO for a large set of models. In this work we have instead implemented the chemical conversion of CO using an approximate description for CO gas and grain-surface chemistry based on the results of \cite{Bosman2018b}. The description simplifies the full chemical network in \cite{Bosman2018b} by only tracing the main carbon carriers, meaning CO, CH$_3$OH, CO$_2$ and CH$_4$. 
The chemistry is split up into the carrier species that have long ($>10^4$ yr) chemical timescales and intermediates, radicals and ions, which have a short, < 100 yr, chemical timescales. The former are explicitly integrated while kinetic equilibrium is assumed for the latter. 
A more detailed description can be found in Appendix A in \cite{Krijt2020}. 

Using this approximate grain-surface chemistry, we recalculate the CO abundances computed by \texttt{DALI}.
In doing so some thought has to be put in where in the disk our method gives an accurate approximation of the full chemical network in \cite{Bosman2018b}. Based on tests where the full chemistry is computed, we identify three boundaries beyond which the shielded midplane approximation is no longer valid: a low outward column of CO, the presence of oxygen in the gas phase, and a high temperature.
The first is an outward column of CO of $N_{\rm CO, outward} = 10^{16}\ \mathrm{cm}^{-2}$ (see also \citealt{vanDishoeckBlack1988,Heays2017}). The outward column is defined as the minimum of the vertical and radial external columns. In regions with a lower outward column CO can be photodissociated by UV photons and the CO abundance is set by the balance between photodissociation and chemistry, which is included in DALI. 

The second boundary is related to the fraction of oxygen in the gas phase. The approximation assumes that all gas-phase oxygen directly transforms into OH ice where it can react with CO or H to form CO$_2$ ice or H$_2$O ice, respectively. The presence of atomic oxygen in the gas phase in the model indicates that photodissociation is still a significant driver of the chemistry. UV dissociation reactions will dissociate carbon carriers other than CO and push carbon back into the more photo-stable CO (see, e.g., \citealt{vanDishoeckBlack1988}). Empirically we have found the boundary where 5\% of the available oxygen exists as gas-phase O or OH encloses the region where in the full chemical network CO is transformed into other species (see Figure \ref{fig: boundary approximation}).

Finally we also exclude the parts of the disk where $T_{\rm gas} > 35\ \mathrm{K}$. Tests show that above this temperature the conversion of CO through grain-surface chemistry is negligible. At these temperatures CO is converted in CH$_4$ in the gas phase (see, e.g., \citealt{Aikawa1997,Schwarz2018}. This is the dominant route above $\sim25$ K and densities below $\lesssim 10^{10}\ \mathrm{cm}^{-3}$. It becomes less and less effective with increasing temperature due to the lower sticking of atomic oxygen on grains 
(see Figure 4 in \citealt{Bosman2018b}). The increased time atomic oxygen spends in the gas phase results in the reformation of CO through reactions of O and OH with hydrocarbon radicals and ions, for example: CH$_2^+$ + O $\rightarrow$ HCO$^+$ $\rightarrow$ CO. 

It should be noted that efficiency of gas-phase conversion of CO depends directly on rate at which CO can be destroyed by He$^{+}$. Changes to the assumptions made to calculate this rate, for example increasing cosmic-ray ionization rate of He, could increase the efficiency of gas-phase CO conversion. 
It should also be noted that our chemical conversion of CO focuses on the reactions that result in CO no longer being the main carbon carrier in the disk. Other reactions that include CO do still occur both inside and outside our set boundaries (e.g., the formation of H$_2$CO; see, e.g., \citealt{Loomis2015,Oberg2017,tvs2020}), but these reactions do not significantly affect CO being the predominant carbon carrier.
To summarize, CO abundances are recalculated using the approximate grain-surface chemistry for regions in the disk where $N_{\rm CO, outward} > 10^{16}$, $\leq 5\%$ of [O] is in O(g)+OH(g) and $T_{\rm gas} \leq 35\ \mathrm{K}$.
The abundances of the CO isotopologs \xco\ and \cyo\ are calculated by scaling their abundances by factor $X_{\rm CO}^{\rm new} / X_{\rm CO}^{\rm old}$, that is, by how much the $^{12}$CO abundance has decreased.

Figure \ref{fig: CO depletion in abundance} shows the effects of CO conversion through grain-surface chemistry on the CO abundances at 1, 3, and 10 Myr for one of our disk models. As discussed above, the CO abundances are only recalculated below the dashed black lines. At 1 Myr the CO abundance is still $\mathrm{CO/H_2}\sim10^{-4}$ and CO has not been depleted significantly. By 3 Myr the CO abundance has dropped to $\mathrm{CO/H_2}\sim5\times10^{-5}$ higher up in the disk and $\mathrm{CO/H_2}\sim5\times10^{-6}$ close to the midplane. 
For \xco\ the effect of isotope-selective photodissociation is partly offset by the reaction $^{13}\mathrm{C}^+$ + $^{12}\mathrm{CO} \rightarrow$ $^{13}\mathrm{CO}$ + $^{12}\mathrm{C}^+ + 35\ \mathrm{K}$, which is energetically favored toward the production of \xco\ at low temperatures (see, e.g., \citealt{Visser2009,Miotello2014} and references therein). 

Having a lower abundance and lacking a back reaction similar to \xco\ (see eq. 2 in \citealt{Miotello2014}), \cyo\ is much more affected by isotope-selective photodissociation. As a result, \cyo\ is confined to a layer deeper in the disk.
This can be seen in Figure \ref{fig: CO depletion in abundance}, with the \xco\ 3-2 emitting region lying higher up in the disk, fully above the black contour. The \cyo\ 3-2 emitting region lies closer toward the midplane and crosses the black contour. We can therefore expect the \xco\ 3-2 flux not to be affected by the chemical conversion of CO, whereas the \cyo\ 3-2 flux will be affected.
After 10 Myr almost all of the available CO has been converted, with $\mathrm{CO/H_2}\lesssim10^{-6}$ for the region where CO can be converted through cosmic-ray-driven chemistry. The \xco\ and \cyo\ emitting regions have moved inward and upward to regions of the disk that still have a high CO abundance.

In this work we raytrace the \xco\ and \cyo\ $J=3\,-\,2$ lines, but tests show that the $J=2\,-\,1$ lines show the same qualitative behavior (cf. Appendices \ref{app: all 3-2 fluxes} and \ref{app: all 2-1 fluxes}). As a rule of thumb if the $J=3\,-\,2$ lines are optically thick, meaning that at early ages, when the disk is most massive, the \xco\ 2-1 line is up to 5\% fainter (in Jy km s$^{-1}$) and the \cyo\ 2-1 line is up to 15\% fainter. If the $J=3\,-\,2$ lines are optically thin, the $J=2\,-\,1$ fluxes are up to 30\% lower for both \xco\ and \cyo.

\section{Results}
\label{sec: results}
\subsection{CO isotopolog line fluxes: A viscously evolving disk}
\label{sec: CO isotopolog line fluxes}

\begin{figure*}
    \centering
    \includegraphics[width=\textwidth]{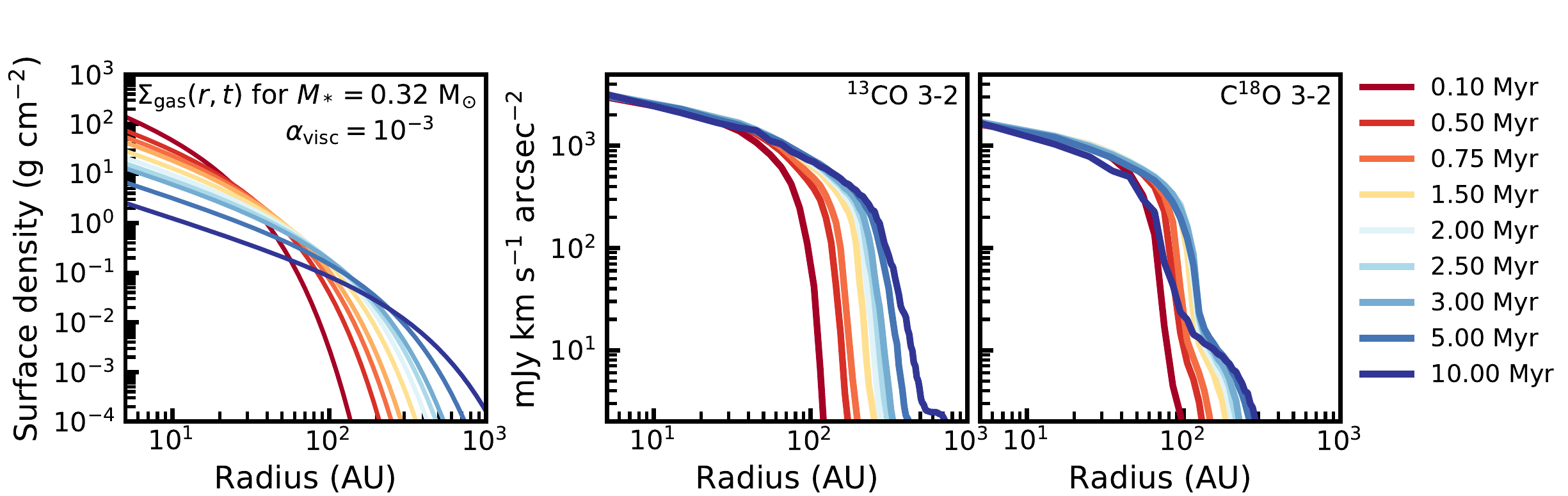}
    \caption{\label{fig: profiles for M032}Time evolution of the surface density profile (left panel), with the middle and right panel showing the corresponding \xco\ $J=3-2$ and \cyo\ $J=3-2$ intensity profiles, respectively. The model shown here has $\mstar=0.32\ \msun, \alpha=10^{-3}$ and $\mdisk(t=0) = 10^{-2}\ \msun$. The colors, going from red to blue, show different time steps. \xco\ and \cyo\ 3-2 intensity profiles for all models can be found in Appendix \ref{app: intensity profiles}}
\end{figure*}

\begin{figure}
    \centering
    \includegraphics[width=\columnwidth]{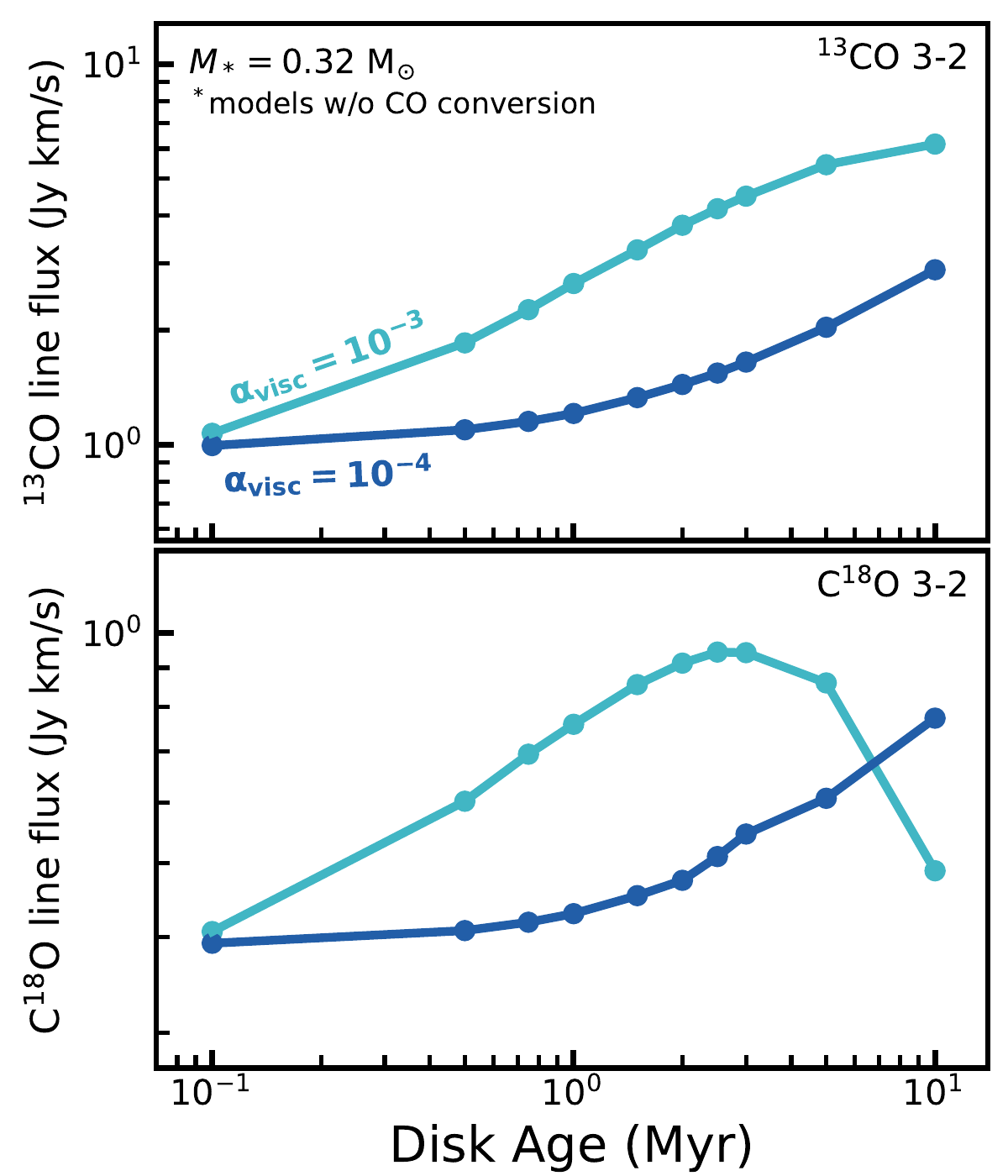}
    \caption{\label{fig: model xCyO fluxes}\xco\ and \cyo\ $J=3-2$ line fluxes of a viscously evolving disk, shown in the top and bottom panel, respectively. Presented here are models with $\mstar=0.32\ \msun$, $\alp=10^{-3}-10^{-4}$ that have an initial disk mass $\mdisk(t=0) = 2\times10^{-3} - 2\times10^{-2}\ \msun$. We note that for these models CO conversion through grain-surface chemistry is not included.}
\end{figure}

As the disk evolves viscously, the characteristic radius increases and the disk spreads out. At the same time the disk mass decreases as material is accreted onto the star. The left panel of Figure \ref{fig: profiles for M032} shows the surface density profile of the model with $\mstar=0.32\ \msun$ and $\alp=10^{-3}$, corresponding to an initial disk mass $\mdisk(t=0)=10^{-2}\ \msun$, as it changes over time.

The middle panel of Figure \ref{fig: profiles for M032} shows the corresponding \xco\ $J=3\,-\,2$ azimuthally averaged intensity profile. Throughout most of the disk the \xco\ emission is optically thick and decreases slowly with radius. In the outer part of the disk, at a radius between $\sim100-400$ AU depending on the age of the model, the \xco\ emission drops off rapidly. Comparing to the left panel in Figure \ref{fig: profiles for M032}, this radius corresponds to a surface density of $\Sigma_{\rm gas} \approx 10^{-3} \mathrm{g\ cm^{-2}}$. Below this density, the \xco\ column is no longer high enough to effectively self-shield and \xco\ is rapidly removed from the disk beyond this radius.

As the \xco\ emission is optically thick throughout almost all of the disk, the integrated \xco\ flux is expected to increase over time as viscous expansion of the disk increases its emitting area. This is indeed what is seen in Figure \ref{fig: model xCyO fluxes}, where the top panel shows the integrated \xco\ 3-2 flux as a function of disk age. \xco\ intensity profiles of all of our models, shown in Appendix \ref{app: intensity profiles}, indicate that the \xco\ emission is optically thick throughout the disk for all disk models with $\alp\leq10^{-3}$.
Overall, the \xco\ flux is therefore more of a disk size tracer than a disk mass tracer.

The rightmost panel of Figure \ref{fig: profiles for M032} shows the time evolution of the \cyo\ intensity profile for the disk model with $\mstar=0.32\ \msun$, $\alp=10^{-3}$. The profile consists of two components: the inner $\sim90$ AU is the emission of the bulk of the \cyo\ in the disk, located just above the midplane. This emission is optically thick up to 3-5 Myr, after which the emission starts to become optically thin.
The second component, visible beyond $\sim100$ AU, is two orders of magnitude fainter than the first component. This emission is optically thin and originates from a lower abundance warm finger of \cyo\ located higher up in the disk (see Figure 2 in \citealt{Miotello2014}).

While the first component of the \cyo\ emission is optically thick, the total \cyo\ increases with age as can be seen in the bottom panel of Figure \ref{fig: model xCyO fluxes}. The total \cyo\ 3-2 flux increases up to $\sim3$ Myr. At this point the \cyo\ emission within $\sim 100$ AU starts to become optically thin and the \cyo\ integrated flux becomes a tracer of the disk mass instead of the disk size. 

Whether the \cyo\ integrated flux is predominantly optically thick and tracing the disk size or optically thin and tracing the disk mass is determined, to first order, by the mass and the size of the disk. For disk ages $\geq1$ Myr the \cyo\ emission of our models is optically thin throughout the disk if $\mstar \leq 0.2\ \msun$ and $\alp\geq10^{-3}$ $(\mdisk(t=1\ \mathrm{Myr}) \leq 10^{-3}\ \msun)$. However, higher mass disks $(\mdisk(t=1\ \mathrm{Myr})$ around higher mass stars $(\mstar\geq0.32\ \msun)$ or disks that have low turbulence $(\alp\leq10^{-4})$ instead have optically thick \cyo\ 3-2 emission throughout most of the disk. 

\subsection{CO isotopolog line fluxes: Effects of grain-surface chemistry}
\label{sec: CO isotopolog line fluxes: effects of grain surface chemistry}

\begin{figure}
    \centering
    \includegraphics[width=\columnwidth]{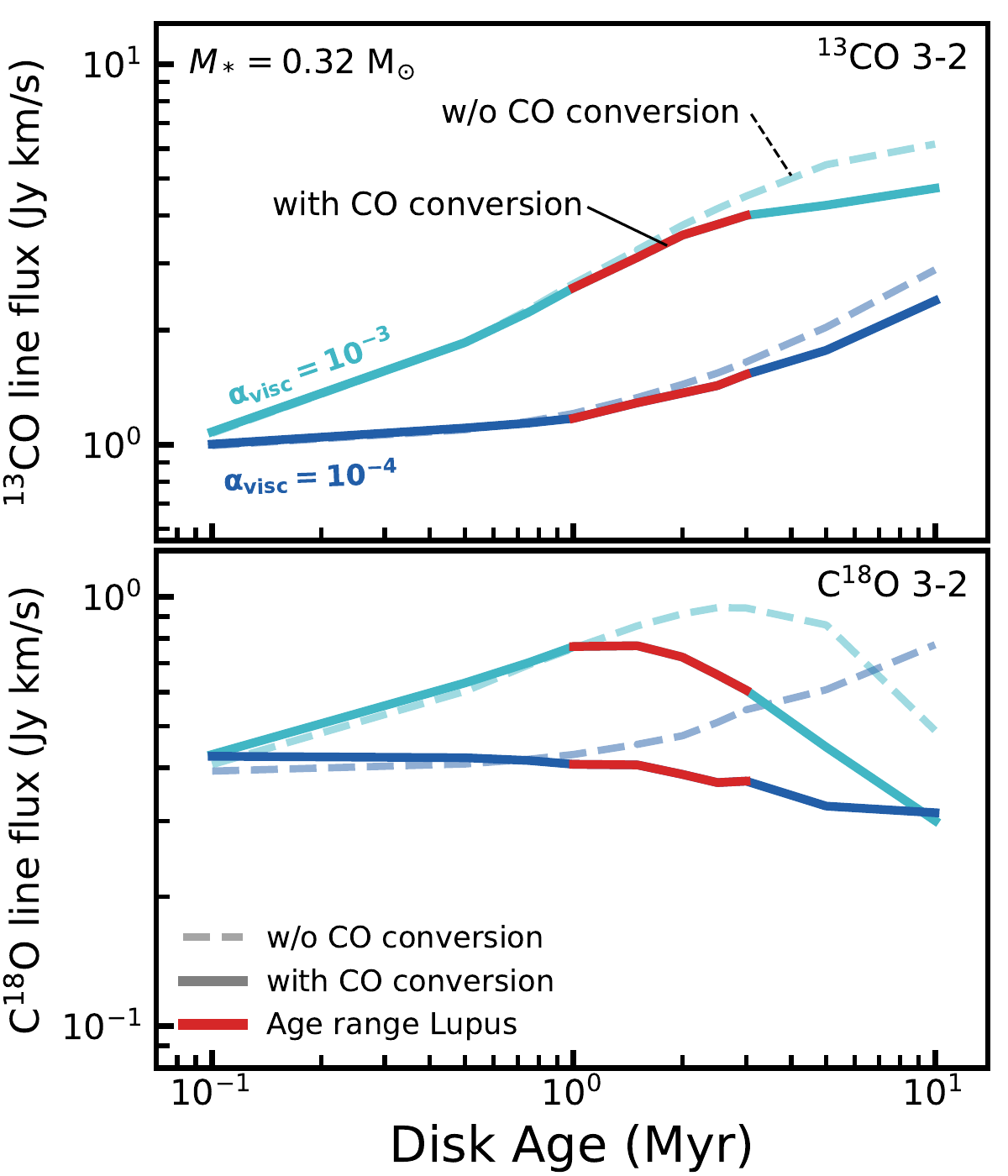}
    \caption{\label{fig: model xCyO fluxes - with gsc}\xco\ and \cyo\ $J=3\,-\,2$ line fluxes of a viscously evolving disk, shown in the top and bottom panel, respectively. Presented here are models with $\mstar=0.32\ \msun$, $\alp=10^{-3}-10^{-4}$ that have an initial disk mass $\mdisk(t=0) = 2\times10^{-3} - 2\times10^{-2}\ \msun$. A full overview of the model fluxes can be found in Figure \ref{fig: all xCyO 3-2 fluxes}. Solid lines show fluxes for models that include CO conversion through grain-surface chemistry. For comparison, dashed lines show fluxes for models without CO conversion as in Figure \ref{fig: model xCyO fluxes}. The red line highlights the age range of disks in the Lupus star-forming region.}
\end{figure}

Before we compare our models to the observations, we first look at the effects of conversion of CO through grain-surface chemistry on CO isotopolog line fluxes. 
As lowering the CO abundance decreases the \xco\ and \cyo\ column densities, we expect the \xco\ and \cyo\ emission to become optically thin at an earlier disk age.

Figure \ref{fig: model xCyO fluxes - with gsc} shows \cyo\ and \xco\ $J=3\,-\,2$ integrated line fluxes for models with $M_* = 0.32\ \mathrm{M_{\odot}}$ and $\alp = [10^{-3},10^{-4}]$. The line fluxes without grain-surface chemistry, presented earlier in Figure \ref{fig: model xCyO fluxes}, are included as dashed lines. Up to 1 Myr there is no discernible difference between the models with and without grain-surface chemistry, as it takes time for the chemistry to lower the gas-phase CO abundance. We note here that the timescale for the grain-surface chemistry depends directly on the assumed cosmic-ray ionization rate \zetacr, which we have assumed to be $\zetacr = 1\times10^{-17}\ \mathrm{s}^{-1}$. For a higher \zetacr\ the conversion of CO into other species will occur on a shorter timescale (see also Section \ref{sec: zeta estimation}). 

For \xco\ there is almost no difference when grain-surface chemistry is included even after 1 Myr. As can be seen in Figure \ref{fig: CO depletion in abundance} the \xco\ 3-2 emitting region lies almost completely above the region of the disk where CO can be transformed on the grains.
Grain-surface chemistry has a larger effect on the \cyo\ fluxes because the \cyo\ 3-2 emitting region lies closer to the midplane, where the conversion of CO is most efficient (see, e.g., Figure \ref{fig: CO depletion in abundance}). For both $\alp = 10^{-3}$ and $10^{-4}$ the \cyo\ 3-2 flux starts to decrease at a disk age of $\sim1.5$ Myr. This indicates that the \cyo\ emission is optically thin for a much larger disk masses (up to $\mdisk \sim 2\times10^{-2}\ \msun$, cf. Figure \ref{fig: disk mass evolution}). At 10 Myr grain-surface chemistry has lowered the \cyo\ 3-2 fluxes by a factor of $\sim2-3$ compared to the model without CO conversion. 

\subsection{Comparing to the Lupus disk population}
\label{sec: comparing to observations}

\begin{figure*}
    \centering
    \begin{subfigure}{0.98\textwidth}
    \includegraphics[width=\textwidth,clip,trim={0cm 1.49cm 0cm 0cm}]{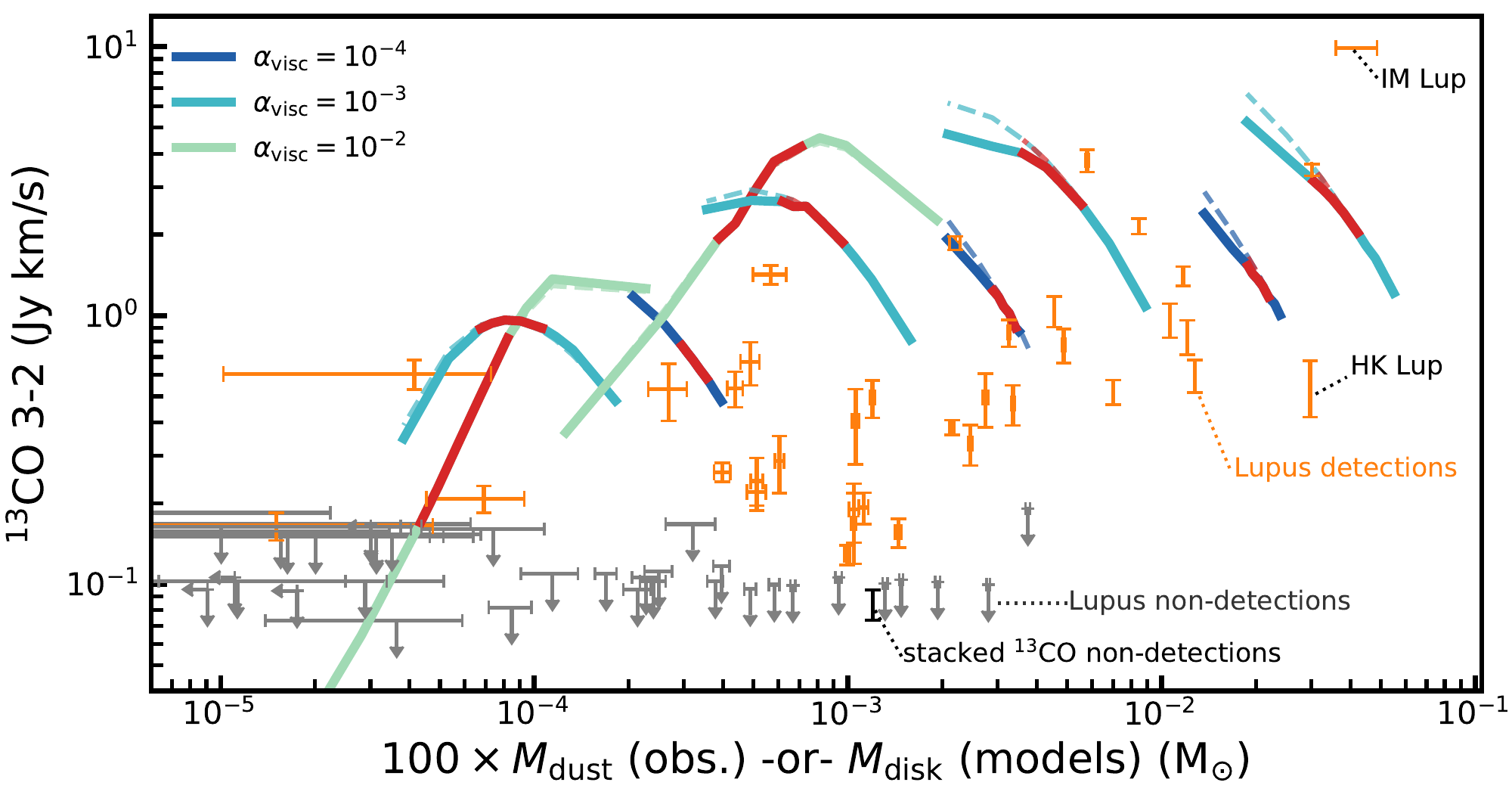}
    \end{subfigure}
    \begin{subfigure}{0.98\textwidth}
    \includegraphics[width=\textwidth]{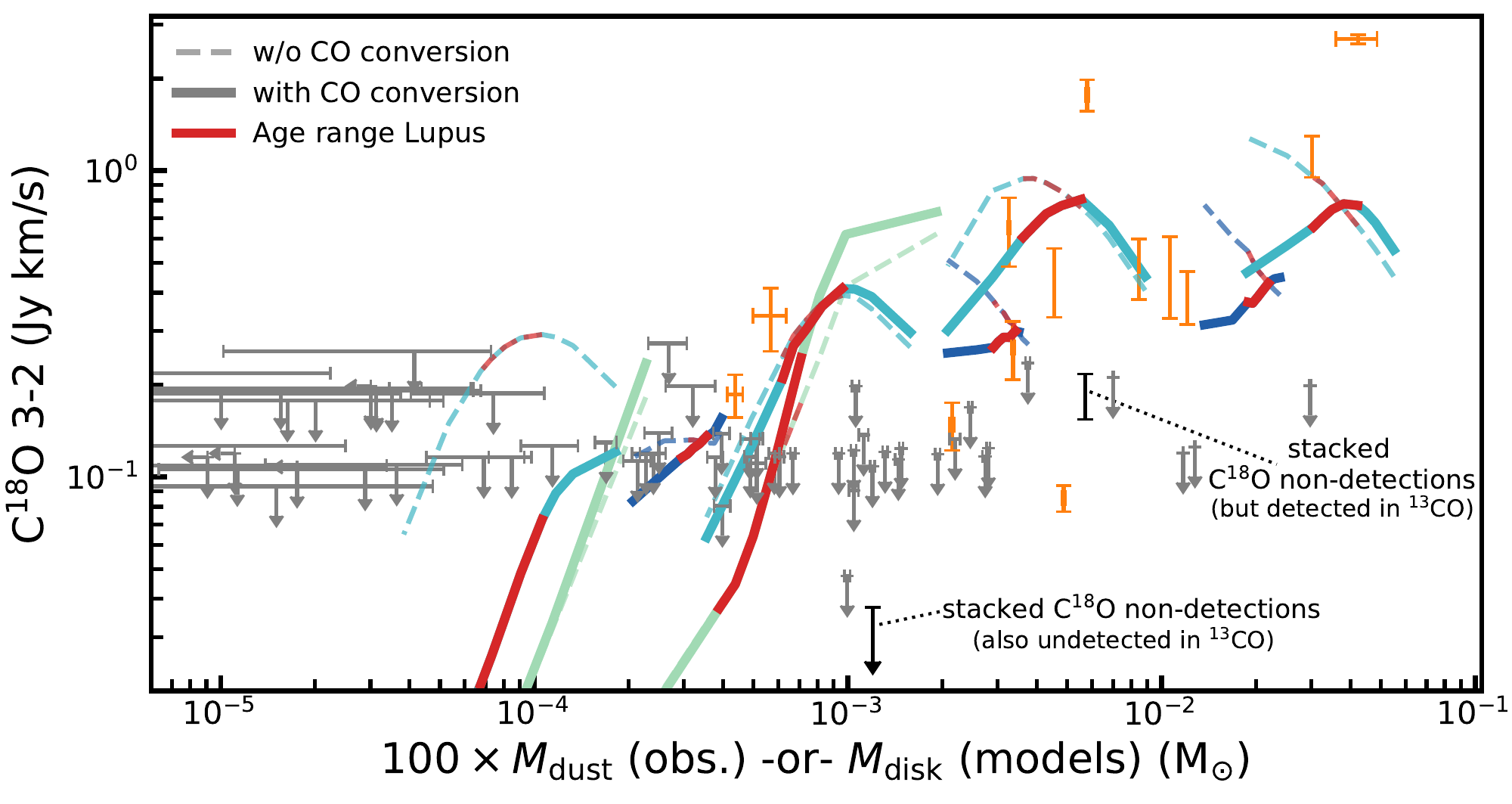}
    \end{subfigure}
    \caption{\label{fig: xCyO fluxes (iso + gsc) vs mdust}
    $^{13}$CO (top) and C$^{18}$O (bottom) $J=3\,-\,2$  line fluxes versus disk mass. 
    Solid lines show models that include CO conversion through grain-surface chemistry. For comparison, dashed lines show the models without grain-surface chemistry.
    For the observations, shown in orange,
    we use $M_{\rm gas} \simeq 100\times M_{\rm dust}$ as a proxy for the disk mass (see, e.g., \citealt{ansdell2016,ansdell2018,Yen2018}). Colors show models with different viscous alpha. Models with age between 1 and 3 Myr, the approximate age of Lupus, are highlighted in red. We note that in our models $M_{\rm disk}$ decreases with time, meaning that time runs right to left in this figure for our models. Observations for which we only have an upper limit on the $^{13}$CO or C$^{18}$O line flux are shown in gray. Stacked non-detections are shown in black (cf. Section \ref{sec: comparing to observations}).
    }
\end{figure*}

Recently \cite{ansdell2016} have carried out an ALMA survey of the protoplanetary disk population in the Lupus star-forming region, in both continuum and $^{12}$CO, \xco\ and \cyo\ line emission (see also \citealt{ansdell2018}). We examine whether the low \xco\ and \cyo\ 3-2 fluxes are compatible with viscous evolution. 

Figure \ref{fig: xCyO fluxes (iso + gsc) vs mdust} shows the \cyo\ and \xco\ $J=3\,-\,2$ integrated line fluxes of our models and observations of protoplanetary disks in Lupus \citep{ansdell2016,Yen2018}.
To have a useful comparison, we have aligned the models and observations based on total (estimated) disk mass. For the observations we use $100\times\mdust$ as a proxy for the disk mass. This is equivalent to assuming that disks have a gas-to-dust mass ratio $\gdrat = 100$, the canonical value for the ISM. 
It is likely that radial drift of larger grains and their subsequent accretion onto the star causes disks to have a $\gdrat > 100$. This would move the observations to the right in Figure \ref{fig: xCyO fluxes (iso + gsc) vs mdust}. We note, however, that substructures detected in the continuum (e.g., \citealt{ALMApartnership2015,Andrews2016,Andrews2018,vanTerwisga2018,Long2018,Long2019,Cieza2020}) could halt the inward drift of large grains (see e.g., \citealt{Pinilla2012,Dullemond2018}). 
This is discussed in more detail in Section \ref{sec: caveats}. In our analysis we exclude transition disks with a resolved inner cavity in the dust continuum as our models do no represent their disk structure (see \citealt{vdMarel2018}). 

At the high mass end $(\mdisk \gtrsim 5\times10^{-3}\ \msun)$ \xco\ 3-2 is detected for all disks and \cyo\ 3-2 is detected for most disks in Lupus.
The observed range of \xco\ fluxes is reasonably well reproduced by our models, although for individual objects the flux for the model with the corresponding mass might be a factor of two to four times higher.
There are two disks, IM Lup (Sz 82) and HK Lup (Sz 98) that are either significantly brighter or fainter than our models. These two disks are among the largest disks in Lupus.
Interestingly, while these two disks are of similar size and dust mass, they differ in \xco\ 3-2 flux by more than an order of magnitude, suggesting that the processes that affect the abundance of CO can be very different in two very similarly looking disks.
We note that our models are aimed at reproducing the average protoplanetary disk and we do not expect them to reproduce outliers. In a similar study \cite{Trapman2020} showed that a larger initial disk size of 30-50 AU is required to reproduce the gas disk size of these large disks. 

The \cyo\ 3-2 fluxes detected for the high mass disks are reproduced by the models within a factor of $\sim2$ for most disks. 
Four of the massive disks in Lupus, slightly less than half of the disks in this mass range, are not detected in \cyo\ 3-2. Our models overproduce these \cyo\ upper limits by a factor of $\sim2-4$, suggesting they either have a $\gdrat \ll 100$ or that they have lower \cyo\ abundances than our models. The \cyo\ fluxes in our models can be reduced by increasing the cosmic-ray ionization rate (see Section \ref{sec: zeta estimation} and Figure \ref{fig: xCyO fluxes (iso + gsc-INF) vs mdust}).

For lower disk masses $(5\times10^{-4}\ \msun \lesssim \mdisk \lesssim 5\times10^{-3}\ \msun)$ the majority of disks in Lupus are still detected in \xco\ but only a few are also detected in \cyo. Here the \xco\ fluxes of our models and the observations start to diverge. While the brightest observed \xco\ fluxes are still reproduced by the models, there is up to an order of magnitude difference between the model \xco\ fluxes and the bulk of the \xco\ detections and the \xco\ upper limits. 

For \cyo, comparing the models to the observations becomes more difficult due to the low number of \cyo\ 3-2 detections in Lupus. The few \cyo\ detections in this mass range are reproduced by the models. However, these are the same disks whose bright \xco\ emission is also reproduced by the models. The \cyo\ fluxes of the models do get low enough to match the observed \cyo\ upper limits $(\sim0.1\ \mathrm{Jy\ km\ s^{-1}})$, but only when the disk models are $\sim10$ Myr old. This is much older than most disks in Lupus, which are estimated to be 1-3 Myr old. 

\cite{ansdell2016} stacked the disks that were detected in the continuum but not in \xco\ 3-2 and \cyo\ 3-2. First, stacking the 25 sources that were detected in the continuum and \xco\ resulted in a mean continuum flux of 70 mJy, corresponding to a dust mass of $\sim5\times10^{-5}\ \msun$, with a detected mean \cyo\ 3-2 flux of $206\pm31\ \mathrm{mJy\ km\ s^{-1}}$. Stacking the 26 sources detected in the continuum but not detected in both \xco\ 3-2 and \cyo\ 3-2 provided a much deeper mean \cyo\ 3-2 upper limit of $42\ \mathrm{mJy\ km\ s^{-1}}$. We note that we have scaled the fluxes to a distance of 160 pc, the average distance to the Lupus clouds based on Gaia DR2 measurements \citep{GaiaDR2_2018,BailerJones2018}, instead of the 200 pc used by \cite{ansdell2016}. This stacked upper limit lies a factor of five to ten lower than our model fluxes, similar to what was found for \xco, suggesting that both \xco\ 3-2 and \cyo\ 3-2 fluxes are overproduced by our models, even if chemical conversion of CO is included.

For disks with $\mdisk=100\times\mdust \lesssim 5\times10^{-4}\ \msun$ only a handful of disks are detected in \xco\ and none are detected in \cyo. While the \xco\ fluxes of the models have decreased for these lower gas masses, the model fluxes are still at least a factor of five higher than the observed \xco\ upper limits. In comparison the \cyo\ fluxes of the model have decreased with \mdisk\ and are consistent with the observed \cyo\ upper limits.

To summarize, our viscously evolving disk models that include CO conversion are able to explain most observed \xco\ and \cyo\ fluxes for high mass disks $(\mdisk \gtrsim 5\times10^{-5}\ \msun)$ in Lupus. For lower mass disks the model \xco\ and \cyo\ fluxes are up to an order of magnitude higher than what has been observed. 

\subsection{Cosmic-ray ionization rate required to match observed CO isotopolog fluxes}
\label{sec: zeta estimation}

\begin{figure}
    \centering
    \begin{subfigure}{0.98\columnwidth}
    \includegraphics[width=\columnwidth,clip,trim={0cm 1.34cm 0cm 0cm}]{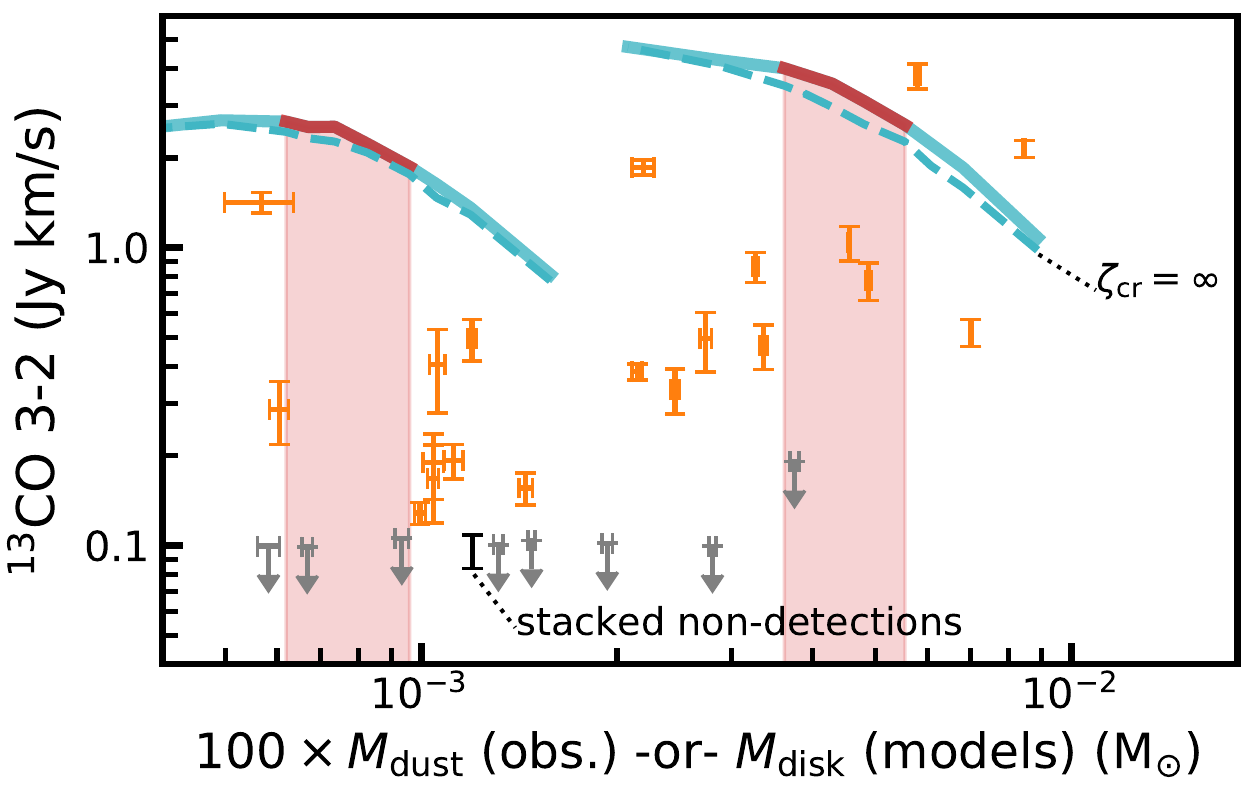}
    \end{subfigure}
    \begin{subfigure}{0.98\columnwidth}
    \includegraphics[width=\columnwidth]{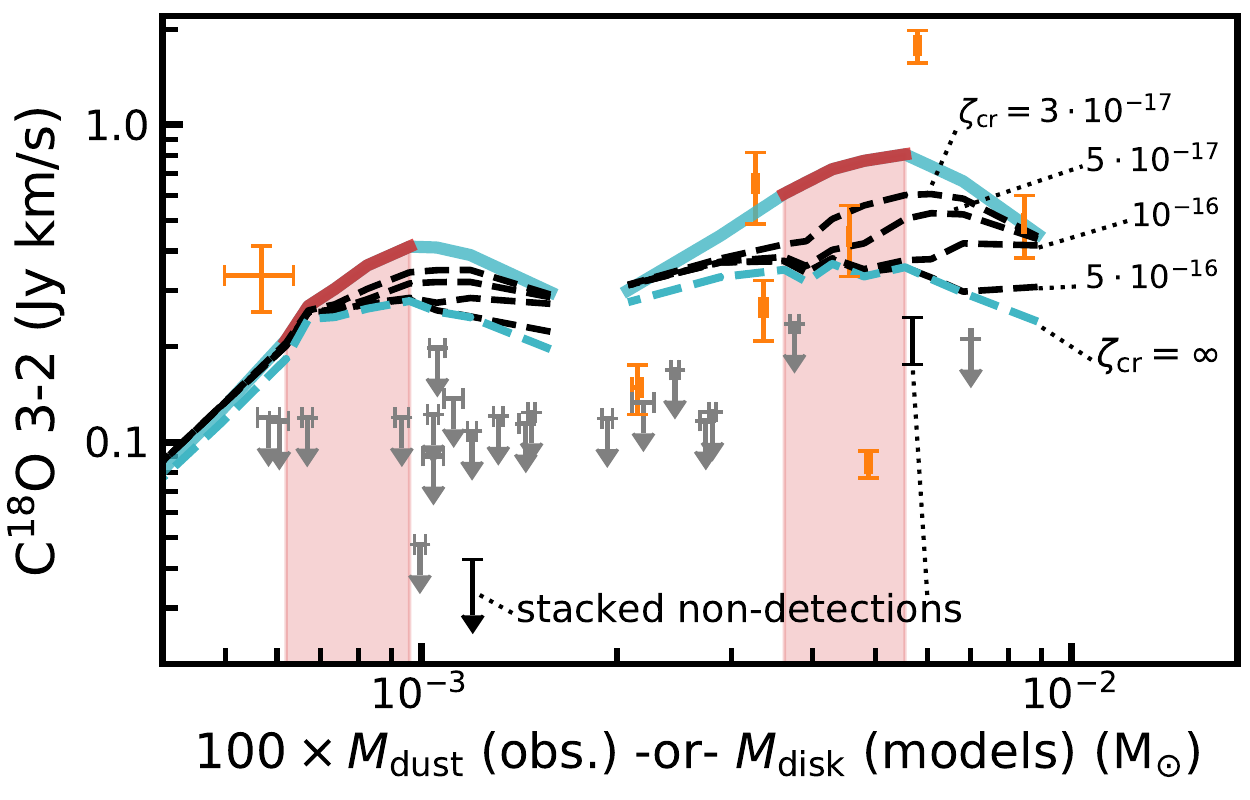}
    \end{subfigure}
    \caption{\label{fig: zeta estimation zoom in}Effect of the cosmic ray ionization rate on \xco\ and \cyo\ $J=3\,-\,2$ line fluxes, shown in the top and bottom panel respectively. Models shown here have $\alp=10^{-3}$ and $\mstar = [\,0.2, 0.32\,]\ \msun$. Solid light blue lines show the model line fluxes where CO conversion through grain-surface chemistry is calculated with a cosmic ray ionization rate $\zetacr = 1\times10^{-17}\ \mathrm{s}^{-1}$. For the black dashed lines \zetacr\ is increased up to $5\times10^{-16}\ \mathrm{s}^{-1}$. The dashed blue line shows the upper limit of the effect of grain-surface chemistry, where we have removed all CO in the region of the disk where CO conversion by grain-surface chemistry is effective. Observations in Lupus are shown in orange if detected and in gray if an upper limit \citep{ansdell2016,Yen2018}. The stacked non-detections are shown in black (see also Section \ref{sec: comparing to observations}). The red shaded lane highlights the age range of the Lupus star-forming region. }
\end{figure}

The previous Section showed that for disk masses below $\sim5\times10^{-3}\ \msun$ the model \xco\ and \cyo\ fluxes are a factor of five to ten times higher than what is observed in Lupus. The cosmic-ray ionization rate \zetacr\ is one of the main factors that determines the rate at which CO is converted into CO$_2$, CH$_4$ and CH$_3$OH. 
In our models we have assumed a moderate $\zetacr\sim10^{-17}\ \mathrm{s}^{-1}$, comparable to the rate found in dense molecular clouds (see, e.g., \citealt{Black1990}). The cosmic-ray ionization rate in disks is highly uncertain, so it is possible that \zetacr\ is higher, especially if activity from the young star at the center of the disk is contributing (see, e.g., \citealt{Rab2017,Padovani2018}).
However, it should be noted that there is also evidence for a much lower cosmic-ray ionization rate in disks $(\zetacr\sim10^{-19}-10^{-18}\ \mathrm{s}^{-1})$ as a result of the disk being shielded by the stellar magnetic field and disk winds \citep{Cleeves2013,Cleeves2015}. However, as reducing \zetacr\ only decreases the effectiveness of the chemical conversion of CO we do not examine it here. Instead we examine if increasing \zetacr\ would allow us to reproduce the observations and what \zetacr\ would be needed to do so in a 1-3 Myr time period.
We focus on the mass range between $\mdisk = 5\times10^{-4}\ \msun$ and $\mdisk = 10^{-2}\ \msun$ where the difference in fluxes between the models and the observations is the largest. Similarly, we focus on models with $\alp =10^{-3}$ and $\mstar = [0.2,0.32]\ \msun$, which span a similar disk mass range.

Figure \ref{fig: zeta estimation zoom in} shows the \xco\ and \cyo\ 3-2 fluxes after CO abundances were recalculated using a higher \zetacr\ than is used in the models presented previously ($\zetacr = 1\times10^{-17}\ \mathrm{s}^{-1}$; see Table \ref{tab: model fixed parameters}.) 
The \xco\ 3-2 emission is only weakly dependent on the cosmic-ray ionization rate, with less than $\sim10\%$ differences in flux. Again, this is due to the \xco\ emitting region $(z/r \sim0.2-0.35)$ being higher than the region of CO removal $(z/r \lesssim0.15)$
In Section \ref{sec: 13co options} we discuss alternative ways to reconcile the \xco\ 3-2 observations with our models.

The bottom panel of Figure \ref{fig: zeta estimation zoom in} shows that, in contrast to \xco, increasing \zetacr\ has a larger effect on the \cyo\ fluxes, but not enough.
Increasing \zetacr\ from $1\times10^{-17}\ \mathrm{s}^{-1}$ to $5\times10^{-17}-10^{-16} \mathrm{s}^{-1}$ is sufficient to explain almost all of the \cyo\ detections for disks with $100\times\mdust \gtrsim 2\times10^{-3}\ \msun$. However, the figure also shows that by increasing \zetacr\ the \cyo\ 3-2 fluxes can reduced to at most $\sim0.2\ \mathrm{Jy\ km\ s^{-1}}$, which is still a factor of $\sim2$ higher than the observed \cyo\ upper limits in Lupus. Moreover, the fluxes obtained from our models remain a factor of $\sim5$ higher than the stacked \cyo\ 3-2 upper limits. This indicates that increasing the amount of cosmic-ray ionization in our models by itself cannot explain the faintest \cyo\ 3-2 fluxes.

\section{Discussion}
\label{sec: discussion}

\subsection{Reproducing \xco\ 3-2 line fluxes observed in Lupus}
\label{sec: 13co options}

In Section \ref{sec: comparing to observations} it was found that our models overproduce the \xco\ 3-2 and \cyo\ 3-2 observation in Lupus by a factor of five to ten for disks with $\mdisk \lesssim 5\times10^{-3}\ \msun$. By increasing the cosmic-ray ionization rate from $\zetacr = 1\times10^{-17}\ \mathrm{s}^{-1}$ to $10^{-16}\ \mathrm{s}^{-1}$ 
it is possible to decrease the \cyo\ 3-2 fluxes of our models to within a factor of two of the observed upper limits.
However, the \xco\ 3-2 emission originates predominantly from a layer higher up in the disk where CO is not being efficiently converted into other species. As such, the \xco\ 3-2 fluxes of our models 
remain a factor of $10-30$ times higher than the observations, even after removing all CO from region of the disk where CO grain-surface chemistry is effective. 
Here, we examine the emitting regions of \xco\ 3-2 and \cyo\ 3-2 and we discuss processes that could reduce the \xco\ and \cyo\ fluxes to the point where they are in agreement with both detections and upper limits seen in observations. 

\begin{figure}
    \centering
        \begin{subfigure}{0.98\columnwidth}
    \includegraphics[width=\columnwidth,clip,trim={0cm 1.14cm 0cm 0cm}]{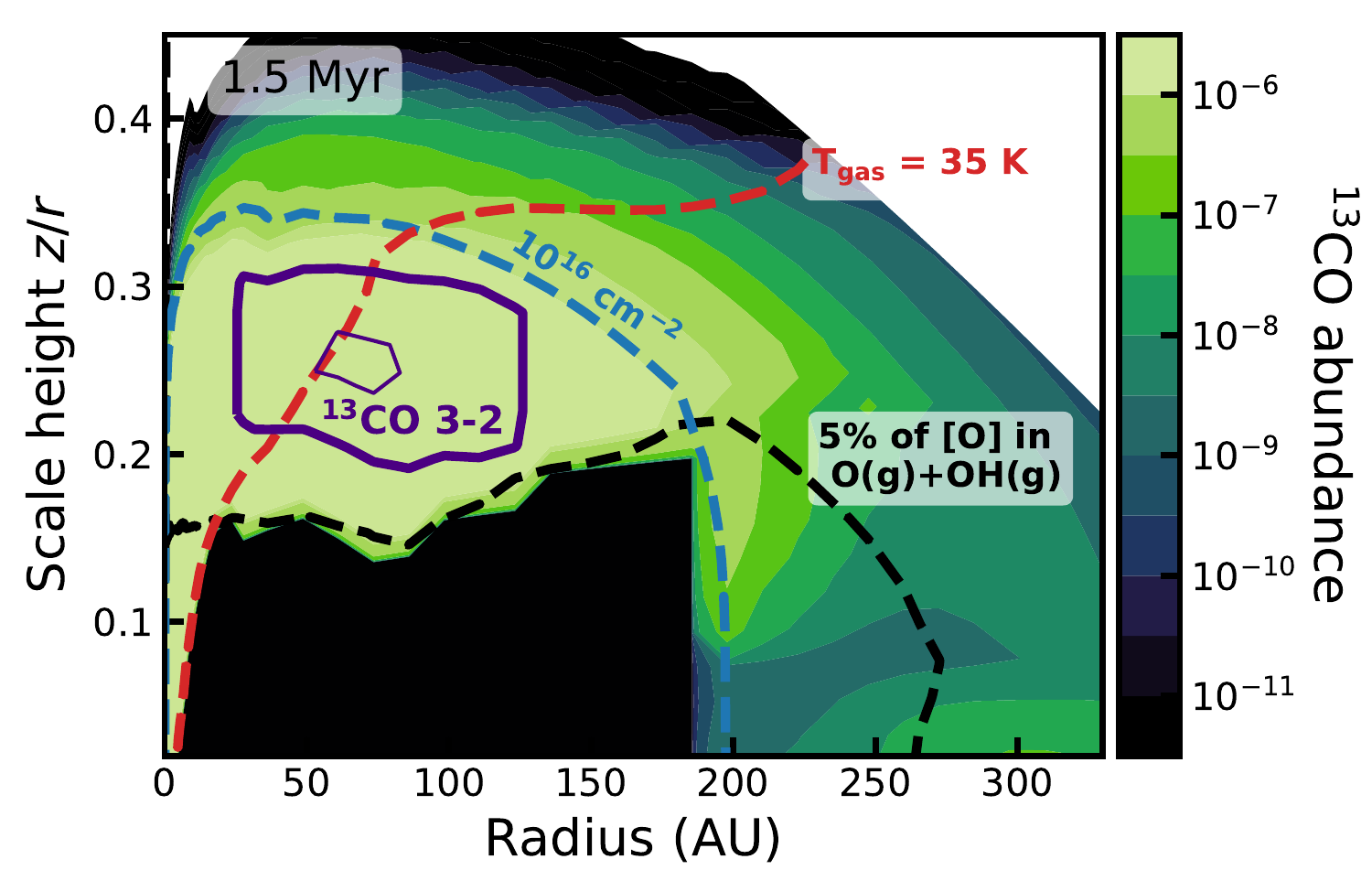}
    \end{subfigure}
    \begin{subfigure}{0.98\columnwidth}
    \includegraphics[width=\columnwidth,clip,trim={0cm 0cm 0cm 0.3cm}]{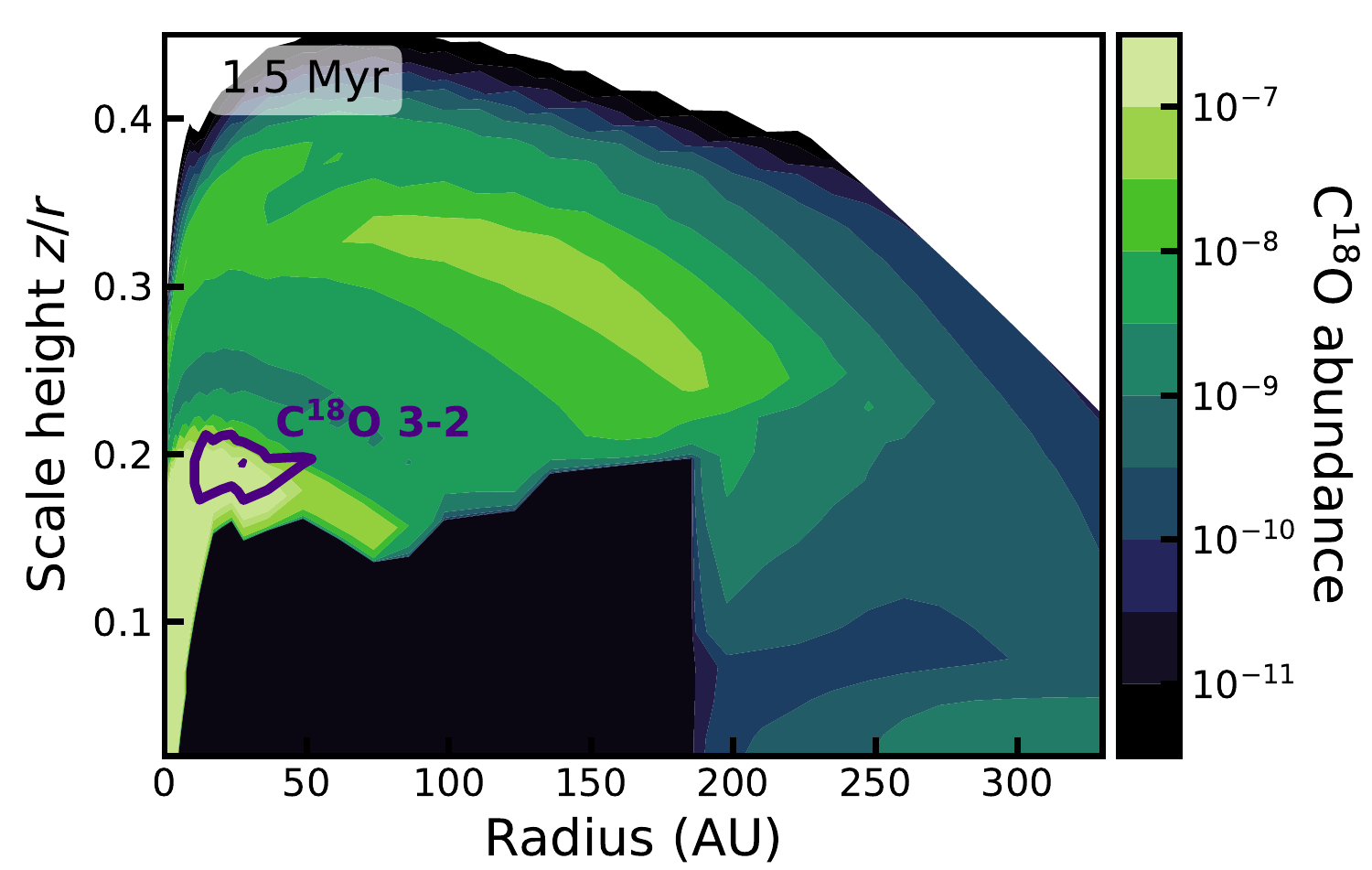}
    \end{subfigure}
    \caption{\label{fig: 13CO emitting layer} $^{13}$CO (top) and C$^{18}$O (bottom) abundance structure (colors) and $J=3-2$ emitting regions (purple contours) for an example disk model ($\mstar = 0.32\ \msun,\ \alp=10^{-3}$ and a disk age of 1.5 Myr). 
    The contours shown are the same as in Figure \ref{fig: boundary approximation}. 
    }
\end{figure}

Figure \ref{fig: 13CO emitting layer} shows an example of the \xco\ and \cyo\ emitting regions in a model with maximum CO conversion. The \xco\ emitting region is located high up in the disk $(z/r \sim0.25-0.4)$, fully above the region where CO can be removed $(z/r\lesssim0.15)$. The presence of oxygen in the gas phase in this layer indicates that it cannot be locked up efficiently in H$_2$O, meaning it is available to react with the carbon released from the destruction of CO to reform CO (see also \citealt{Schwarz2018}). It should be noted a small amount of carbon can go into other molecules such as H$_2$CO (e.g., \citealt{Loomis2015,Oberg2017,tvs2020}) and C$_2$H (e.g., \citealt{Kastner2015,Bergin2016,Cleeves2018,Bergner2019,Miotello2019}), but in this FUV dominated layer CO remains the main carbon carrier (see Figure 18 in \citealt{Schwarz2018}).
The \cyo\ emitting region, shown in the bottom panel of Figure \ref{fig: 13CO emitting layer} lies much deeper in the disk, at a height of $z/r\sim0.18$. With efficient CO conversion the \cyo\ emitting region is very compact and extends only slightly above the oxygen-threshold. 
Unlike for \xco\ it is therefore plausible that a slight change in for example the height of the disk will push the \cyo\ abundant region below the oxygen-threshold, thus drastically reducing the observed \cyo\ 3-2 fluxes. In the rest of this section we therefore focus on lowering the \xco\ 3-2 fluxes, which are in stronger violation with the observations.

Knowing the location of the \xco\ emission, we can discuss processes by which the emission can be reduced.

\emph{A colder disk:}
Given that the \xco\ 3-2 emission is optically thick, a lower disk temperature could potentially explain the low \xco\ 3-2 emission in Lupus. Lowering $T_{\rm gas}$ would decrease the brightness of the optically thick line. 
The temperature of a disk is predominantly set by its vertical structure. Our models all have the same height, $H = Rh = h_c\left(R/R_c\right)^{\psi} = 0.1 \left(R/R_c\right)^{0.15}$ (see Section \ref{sec: DALI models}). In reality, however, it is unlikely that all disks have the same height. The vertical structure that is assumed in our disk models is thought to be a good representation of the average disk vertical structure, but there are only a few disks, biased toward high disk masses, for which the disk height has been estimated using the scattering surface observed in scattered light (see, e.g., \citealt{Avenhaus2018,Garufi2020}). It is therefore possible that lower mass disks are much flatter $(h_c \ll 0.1)$, and therefore colder, than previously thought. 

A lower scale height would also increase the fraction of CO in the disk that can be converted.
Reducing the scale height of the disk would concentrate the mass in the shielded midplane, where the cosmic-ray-driven conversion of CO is efficient.

A colder disk cannot explain all of the disks observed to be underabundant in CO. The TW Hya disk is thought to be underabundant in CO by a factor of 10-100 (see, e.g., \citealt{Favre2013,Du2015,Bergin2016,Kama2016,Trapman2017,McClure2020}). This disk is also known to be much more flared $(\psi\sim0.3)$, and therefore warmer, than the disk models presented here (see, e.g., \citealt{Kama2016,vanBoekel2017,Schwarz2016,Calahan2020}). Although anecdotal, the example of TW Hya shows that a colder disk cannot be the sole explanation for the low observed \xco\ 3-2 fluxes (see also \citealt{Fedele2016}). 

\begin{figure}
    \centering
    \centering
    \begin{subfigure}{0.98\columnwidth}
    \includegraphics[width=\columnwidth,clip,trim={0cm 1.34cm 0cm 0cm}]{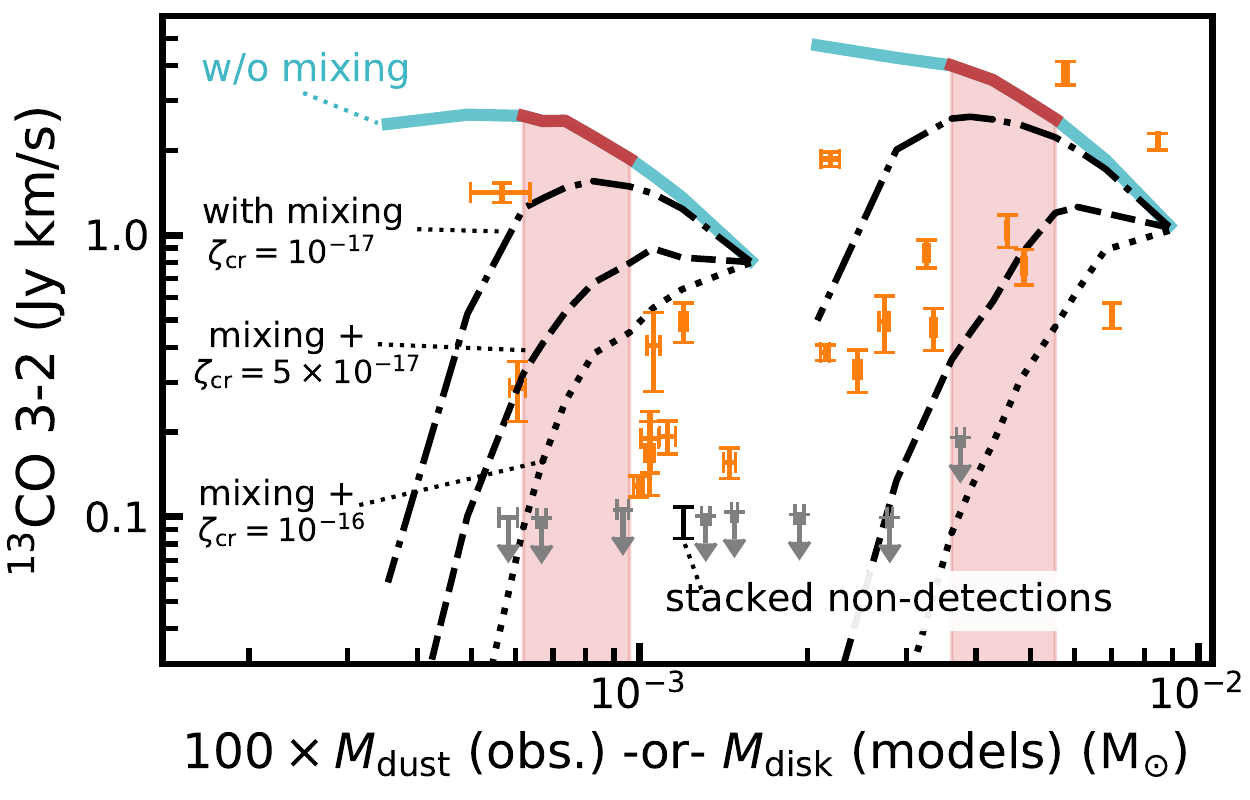}
    \end{subfigure}
    \begin{subfigure}{0.98\columnwidth}
    \includegraphics[width=\columnwidth]{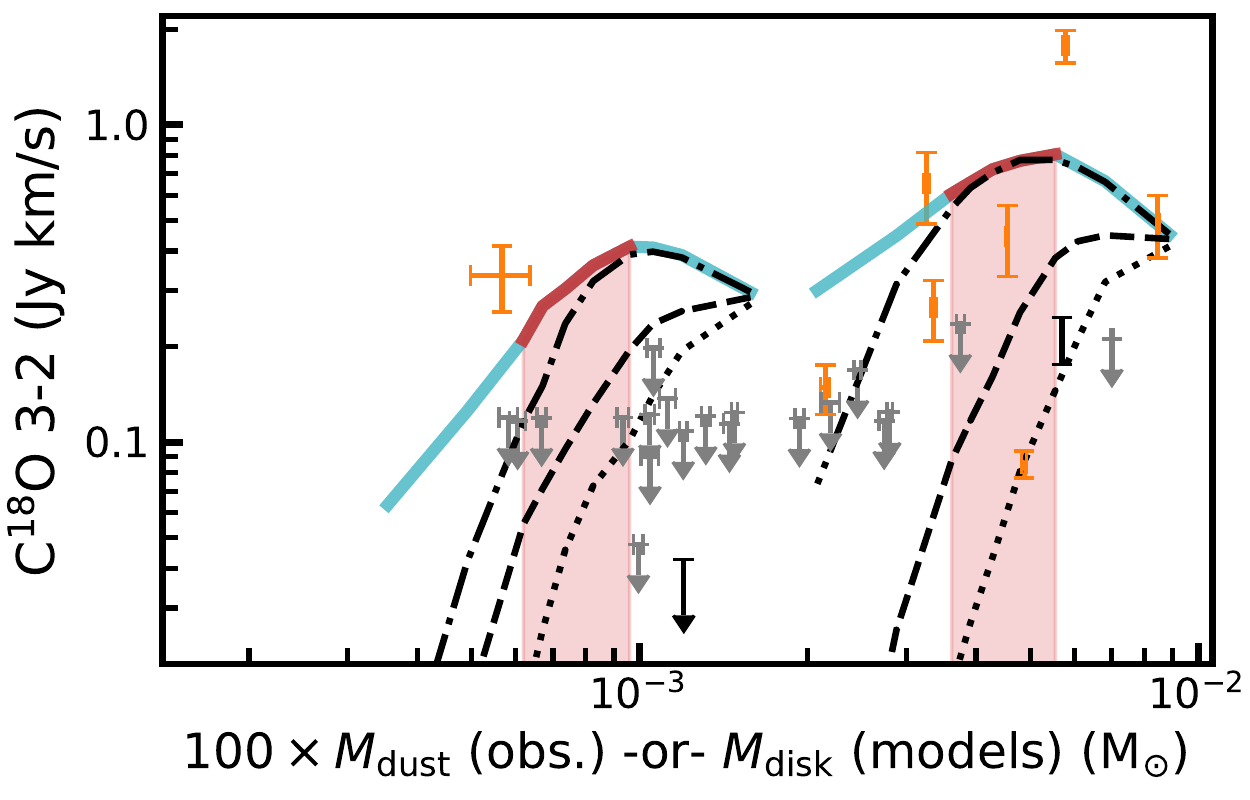}
    \end{subfigure}
    \caption{\label{fig: effect of mixing} \xco\ 3-2 model fluxes after applying vertical mixing (dash-dotted black lines) compared the observation in Lupus (orange markers). Upper limits are shown in gray. The stacked non-detections are shown in black (see also Section \ref{sec: comparing to observations}). Shown here are disk models with $\alp=10^{-3}$ and $\mstar = [\,0.2, 0.32\,]\ \msun$. For reference, the model fluxes without vertical mixing are shown in light blue. The dashed and dotted black lines show models where mixing is combined with higher cosmic-ray ionization rates (see also Section \ref{sec: zeta estimation}). The red shaded lane highlights the age range of the Lupus star-forming region. }
\end{figure}

\emph{Smaller disks:}
Another way to explain the low \xco\ 3-2 fluxes would be that disks are smaller than our models. 
\cite{Trapman2020} showed that our models are consistent with observed gas disk sizes in Lupus, measured from the extent of the $^{12}$CO $J=2\,-\,1$ emission. However, these observations are biased toward the most massive disks around the most massive stars in the Lupus disk population. Comparing their results with the observations presented in Figure \ref{fig: xCyO fluxes (iso + gsc) vs mdust} reveals that there are only three disks below $\mdisk = 3\times10^{-3}\ \msun$ for which the gas disk size has been measured: Sz~133 (238 AU), Sz~65 (172 AU), and Sz~73 (103 AU). Of these, only Sz~73 has a \xco\ flux that lies a factor of $\sim2$ below the predictions of our models. For the remaining disks of similar or lower mass the gas disk size is unknown. Without knowing their gas disk size, it could be possible that the low \xco\ fluxes can be explained by a compact gas disk.

The optically thick emitting region of our disk models with $\mdisk \lesssim 5\times10^{-3}\ \msun$ has a radius of $\sim30-150$ AU, depending on the mass and \rc\ of the disk model. The \xco\ 3-2 flux scales with emitting area $(F\propto R^2)$.
To reduce the \xco\ 3-2 flux of these models by a factor of $10-30$ and bring them in line with the observations, the disk size has to be reduced by a factor of three to five, to $\sim10-50$ AU. We note that this is the disk size as measured from the \xco\ 3-2 emission. \cite{Trapman2019a} found that the gas disk size measured from $^{12}$CO is 30-35\% larger than the gas disk size measured from \xco, assuming that both $^{12}$CO and \xco\ are optically thick. This would suggest that disks with dust masses $\lesssim3\times10^{-4}\ \msun$ need to have gas disk sizes on the order of $\sim40-70$ AU as measured from $^{12}$CO if small disks are the explanation for the low \xco\ fluxes.
It should be noted however that disks with similar \xco\ flux can have very different gas disk sizes (e.g., Sz~129 (140 AU) and HK~Lup (358 AU)), suggesting that \xco\ might not always be optically thick and that compact disks might not be the sole explanation for the faint \xco\ emission seen in observations.
Deeper high resolution $(\lesssim0\farcs2)$ $^{12}$CO and \xco\ observations of low mass disks are required to test this theory.

\emph{Vertical mixing:}
It is very likely that turbulence in protoplanetary disks mixes material both vertically and radially. Through vertical mixing the CO-rich material higher up in the disk would be moved down toward the midplane, where the CO can then be converted into other species via grain-surface chemistry. 
If the mixing timescale is much shorter than the lifetime of the disk the CO will be well mixed and the CO abundance higher up in the disk will match that of the CO-poor material close to the midplane (see, e.g., \citealt{Willacy2006,SemenovWiebe2011,Krijt2018,Krijt2020}).

Figure \ref{fig: effect of mixing} shows the effect of vertical mixing on the \xco\ and \cyo\ 3-2 fluxes. Here we have approximated the effect of mixing by setting the CO abundance higher up in the disk equal to the mass-weighted average CO abundance in the region where we compute the grain-surface chemistry. 
To obtain such efficient mixing requires a mixing timescale that is much shorter than the disk age.
Based on turbulent diffusion, \cite{Krijt2020} calculate a vertical mixing timescale of $t_z \approx 2.5\times10^{5}$ yr for one scale height at 30 AU assuming a turbulent strength corresponding to $\alpha = 10^{-4}$, which is also the lowest level of turbulence assumed in our models. 
We have also assumed that mixing goes up to the \xco\ emitting layer at $z/r\sim0.25-0.4$. 

Up to 3 Myr vertical mixing does not significantly affect the \xco\ and \cyo\ fluxes for the standard case of $\zetacr = 1\times10^{-17}\ \mathrm{s}^{-1}$. While the material in the upper layer of the disk is now well mixed with the material below, the grain-surface chemistry has not had enough time to decrease the mass-averaged CO abundance to the point where it has a noticeable effect on the isotopolog line fluxes. After 3 Myr both the \xco\ flux has started to decrease rapidly and by 10 Myr the fluxes from the our models are at or below the level of the observed upper limits in Lupus.

The effect of mixing on the \xco\ and \cyo\ 3-2 fluxes is limited to how quickly the chemistry is able to decrease the mass-averaged CO abundance. It is therefore interesting to examine whether a combination of vertical mixing and a higher cosmic-ray ionization rate \zetacr\ can explain the \xco\ 3-2 observations in Lupus within 1-3 Myr age of the region.
We included vertical mixing in our models with $\zetacr = 5\times10^{-17}-10^{-16}\ \mathrm{s^{-1}}$ and present the resulting fluxes in Figure \ref{fig: effect of mixing}. With CO conversion now occurring at an increased rate, vertical mixing has a much larger impact at earlier disk ages. Within 3 Myr the model fluxes have decreased to the same level as the faintest \xco\ and \cyo\ detections and upper limits in Lupus.
By combining vertical mixing with a high cosmic-ray ionization rate of $\zetacr = 5\times10^{-17}-10^{-16}\ \mathrm{s}^{-1}$ we can explain all \xco\ 3-2 and \cyo\ 3-2 observations in Lupus. 

This is in agreement with recent results of \cite{Krijt2020}, who investigated processes that have been invoked to explain low CO abundances inferred for disks, such as chemical conversion of CO, turbulent diffusion of gas and small grains and locking up CO in large bodies. 
They show that a combination of vertical mixing and the chemical conversion of CO can decrease the CO abundance to $10^{-6} - 10^{-5}$ up to $z/r\sim0.2$, which is the upper edge of their model. This is similar to the mass-weighted average CO abundance of the viscous disk models that reproduce the observed \xco\ and \cyo\ 3-2 upper limits (see Figure \ref{fig: effect of mixing}). We should note however that to lower the \xco\ flux mixing must be efficient up to at least $z/r\sim0.3$ (see Figure \ref{fig: 13CO emitting layer}).
\cite{Krijt2020} also show that including the locking up of CO ice into larger bodies in addition to the chemical conversion of CO can further decrease the CO abundance to below $10^{-6}$ at $z/r\sim0.2$ for $\zetacr = 1\times10^{-17}\ \mathrm{s}^{-1}$. The combination of these processes could therefore potentially explain the low observed fluxes without having to invoke a high cosmic-ray ionization.

The effect of vertical mixing combined with the chemical conversion of CO is expected to lower the total volatile carbon abundance in the upper layers of the disk. Observations of atomic carbon in TW Hya indeed show evidence that the upper layer of the disk is carbon poor (see \citealt{Kama2016}), but there is insufficient observational evidence to show that this extends to all protoplanetary disks.
\cite{Krijt2020} also find that vertical mixing and chemical conversion of CO leads to (C/O)$_{\rm gas}>1$ in the upper layer of the disk (see their Figure 10). This prediction is in agreement with the bright ring-shaped C$_2$H emission that has been detected in a number of disks (see, e.g., \citealt{Kastner2015,Bergin2016,Cleeves2018,Miotello2019,Bergner2019}). However, it should be noted that it is still unclear how the C$_2$H emission is related to the total amount of carbon that has been removed from the disk (see e.g., the discussion in \citealt{Miotello2019}).

\subsection{Alternative explanations}
\label{sec: caveats}

\emph{Assumed gas-to-dust mass ratio:}
In Section \ref{sec: comparing to observations} we compared our models to observations of disks in Lupus. In this comparison we used $\mdisk \approx 100\times\mdust$, which is equivalent to assuming that the disk has inherited the ISM gas-to-dust mass ratio of $\gdrat = 100$. However, most of the dust mass in protoplanetary is made up of large grains, which are expected to radially drift inward where they are accreted onto the star. This process would lead to $\gdrat \gg 100$, with simulations of dust evolution in protoplanetary disks showing values as high as $\gdrat =  10^{3}-10^{4}$ (see, e.g., \citealt{Birnstiel2012}). 
Increasing the assumed gas-to-dust mass ratio would shift the observations to the right in Figure \ref{fig: xCyO fluxes (iso + gsc) vs mdust}, as it would increase the gas mass we associate with each of the sources. While this could change which models the observations are compared to, it would not significantly affect our results. A small increase $(\gdrat = 200-300)$ might help the comparison between the observations and our models at the high mass end. The main effect of increasing \gdrat\ would be increasing the gas mass threshold, currently $\mdisk \lesssim 5\times10^{-3}\ \msun$, below which we require a higher cosmic-ray ionization rate to match our models to the observations (see also Figure \ref{fig: xCyO fluxes (iso + gsc-INF) vs mdust}).

We note here that lowering the gas-to-dust mass ratio of the observed sources to $\gdrat = 1-10$ would also allow us to reproduce the observed \xco\ and \cyo\ fluxes with our models. This is in essence the same result as was obtained by \cite{ansdell2016} and \cite{miotello2017}, who showed that gas masses derived from the \xco\ and \cyo\ fluxes suggest that disks have low gas-to-dust mass ratios. However, these lower gas masses would no longer be consistent with observed stellar mass accretion rates under the assumption of viscous evolution (see \citealt{Manara2016}).

On the other side of this comparison are the initial disk masses used for our models. As outlined in Section \ref{sec: initial condition}, the initial disk mass $\mdisk(t=0)$ is set by the stellar accretion rate \macc, for which we took the representative \macc\ for four stellar masses, $\mstar = [\,0.1, 0.2, 0.32, 1.0\,]\ \msun$, from observations. 
As the observations show a spread in \macc\ there should be a similar spread in $\mdisk(t=0)$. 
Changing the initial disk mass would, to first order, move the model curves to the left or right in Figure \ref{fig: xCyO fluxes (iso + gsc) vs mdust}.  
While the gas masses of both the models and observations can be varied to some degree, changing the gas masses alone cannot explain the order of magnitude difference between the \xco\ fluxes of models and the observations.

\emph{The assumption of viscous evolution:}
Throughout this work we have assumed that protoplanetary disks evolve viscously and calculated the time evolution of \xco\ and \cyo\ 3-2 line fluxes based on this premise. However, it is also possible disk evolution is instead driven by magnetic disk winds. 
While a quantitative analysis of this scenario is beyond the scope of this work, we can discuss the expected differences. 
In our models we saw that the \xco\ fluxes, and to some degree the \cyo\ fluxes, increase with time while the disk mass instead decreases with time. This is attributed to a combination of the \xco\ emission being mostly optically thick and the disk viscously spreading as it evolves. If disk evolution is instead driven by disk winds the disk is not expected to spread out and the \xco\ and \cyo\ line fluxes will not increase over time. 
Instead, the \xco\ and \cyo\ fluxes are expected to remain constant with time while the emission is optically thick. 
To drive the observed stellar mass accretion the disk mass has to decrease over time. At some point the \xco\ and \cyo\ emission will become optically thin and the line fluxes will start to decrease over time, similarly to what is seen for \cyo\ at 2 Myr in Figure \ref{fig: model xCyO fluxes}. It might take more time for the emission to become optically thin because, in contrast to a viscously spreading disk, the disk mass is not distributed out over an increasingly larger area. Indeed, if disks start out small as suggested by observations (e.g \citealt{Maury2019,Tobin2020}), it is possible that the \xco\ emission remains optically thick as most of the mass is concentrated in a small area.
To test whether disk evolution is driven by disk winds or by viscous spreading, future observations should focus on detecting and resolving the $^{12}$CO, \xco\ and \cyo\ emission of the disks that make up the bulk of the disk population of star-forming regions spread over a wide age range.

\section{Conclusions}
\label{sec: conclusions}

In this work we have used the thermochemical code \texttt{DALI} to run a series of viscously evolving disk models with initial disk masses based on observed stellar mass accretion rates. Using these models we examined how CO isotopolog line fluxes, commonly used as to measure disk gas masses, change over time in a viscously evolving disk. We also compared our models to \xco\ $J=3\,-\,2$ and \cyo\ $J=3\,-\,2$ observations of disks in the Lupus star-forming region, to investigate if they are consistent with disks evolving viscously.
Here we present our conclusions:

\begin{itemize}
    \item \xco\ and \cyo\ 3-2 fluxes of viscously evolving disks increase over time due to the lines being optically thick and their optically thick emitting area increasing in size as the disk expands. 
    Only for disks around stars with $\mstar \leq 0.2\ \msun$ that evolve with a moderate amount of turbulence $(\alp\geq10^{-3})$ the \cyo\ 3-2 emission is optically thin throughout the disk and the integrated \cyo\ flux thus trace the disk mass.
    
    \item Including the conversion of CO through grain-surface chemistry at $< 35$ K does not affect the \xco\ flux. Initially the \cyo\ is also not affected, but from $\sim1$ Myr onward it starts to decrease up to a factor of $\sim2-3$ at 10 Myr. This also ensures that from $\sim1$ Myr and onward \cyo\ 3-2 emission decreases with time in a viscously evolving disk.
    
    \item The observed \xco\ 3-2 and \cyo\ 3-2 line fluxes of the most massive disks $(\mdisk \gtrsim5\times10^{-3}\ \msun)$ in Lupus are consistent to within a factor of two with our viscously evolving disk models where CO is converted into other species through grain-surface chemistry, assuming a moderate cosmic-ray ionization rate $\zetacr\sim 10^{-17}\ \mathrm{s}^{-1}$.
    
    \item Increasing the cosmic-ray ionization rate to $\zetacr\gtrsim 5\times10^{-17}-10^{-16}\ \mathrm{s}^{-1}$ decreases the \cyo\ fluxes to within a factor of $\sim2$ of the observed upper limits for disks in Lupus with $\mdisk \lesssim5\times10^{-3}\ \msun$. Reproducing the stacked \cyo\ upper limit observed in Lupus requires a lower average abundance, which could be obtained with efficient vertical mixing.  
    
    \item Our models overpredict the observed \xco\ 3-2 fluxes by a factor of $10-30$ for most disks with $\mdisk \lesssim 5\times10^{-3}\ \msun$ because the \xco\ 3-2 emission originates from a layer at $z/r\sim0.25-0.4$ ($=2.5-4\times h_c$ for our models), which is much higher up than the region where CO can be efficiently converted into other species $(z/r \lesssim0.15)$. 
    
    \item Reproducing the \xco\ 3-2 observations in Lupus requires both efficient vertical mixing and a higher cosmic-ray ionization rate, $\zetacr\sim 5\times10^{-17}-10^{-16}\ \mathrm{s}^{-1}$. Alternatively, the observations can be explained if less massive ($\mdust \lesssim3\times10^{-5}\ \msun$) disks are either much flatter and colder or much smaller $(\rgas\sim40-70\ \mathrm{AU})$ than their more massive counterparts.
    
\end{itemize}

Our models show that the observed \cyo\ fluxes in Lupus are consistent with these disks having evolved viscously, if CO has been converted into other species under a high cosmic-ray ionization rate. The observed \xco\ fluxes are also consistent with this picture, provided that the material in the disk is also well mixed vertically. However, alternative explanations for the low observed \xco\ fluxes such as the disks being colder or smaller than assumed cannot be discarded based on current observations. Deeper observations that resolve the CO isotopolog emission of low mass disks are needed to conclusively demonstrate whether these disks are evolving viscously. 
Observing the products into which CO is being converted will be difficult as they are either frozen out or their emission lines lie in the infrared. However, with the James Webb Space Telescope it will be possible to search for CO$_2$ and CH$_3$OH ice at intermediate disk layers of edge-on disks or for CO$_2$ gas in the inner disk (see, e.g., \cite{Bosman2017,Anderson2021} for a more detailed discussion).

\begin{acknowledgements}
We thank the referee for the useful comments that helped improve the manuscript.
LT and MRH are supported by NWO grant 614.001.352. ADB acknowledges support from NSF Grant\#1907653  and NASA grant XRP 80NSSC20K0259. GR acknowledges support from the Netherlands Organisation for Scientific Research (NWO, program number 016.Veni.192.233). GR also acknowledges an STFC Ernest Rutherford Fellowship (grant number ST/T003855/1). Astrochemistry in Leiden is supported by the Netherlands Research School for Astronomy(NOVA). All figures were generated with the \texttt{PYTHON}-based package \texttt{MATPLOTLIB} \citep{Hunter2007}. This research made use of Astropy,\footnote{http://www.astropy.org} a community-developed core Python package for Astronomy \citep{astropy:2013, astropy:2018}.
\end{acknowledgements}

\bibliographystyle{aa}
\bibliography{references}

\begin{appendix}

\section{Model \xco\ and \cyo\ $J=3\,-\,2$ intensity profiles}
\label{app: intensity profiles}

\begin{figure*}
    \centering
    \includegraphics[width=\textwidth]{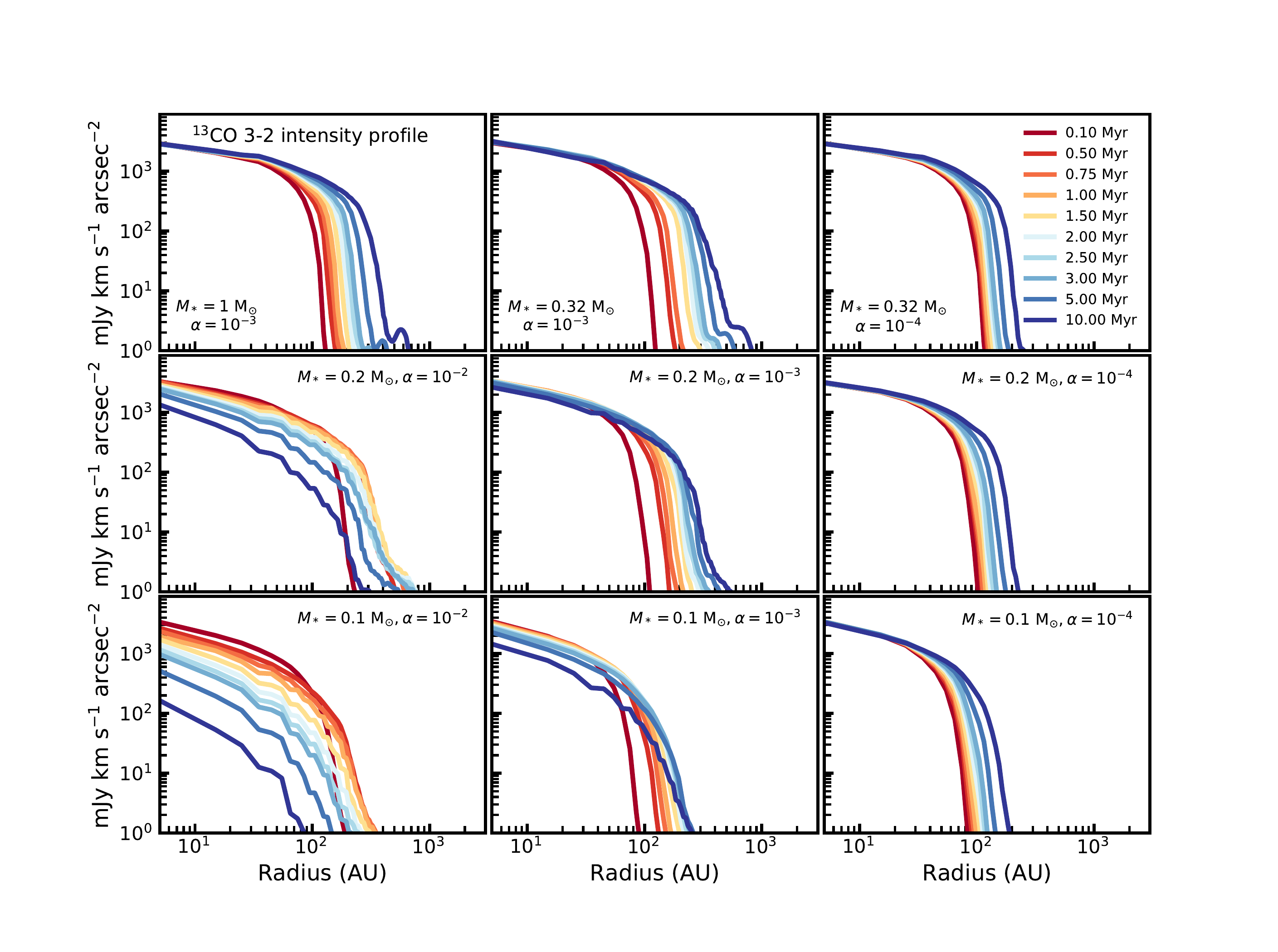}
    \caption{\label{fig: 13CO profiles}Time evolution of the \xco\ $J=3\,-\,2$ intensity profiles of our models, where colors denote different time steps. Each panel shows a different \mstar\ and \alp.}
\end{figure*}

\begin{figure*}
    \centering
    \includegraphics[width=\textwidth]{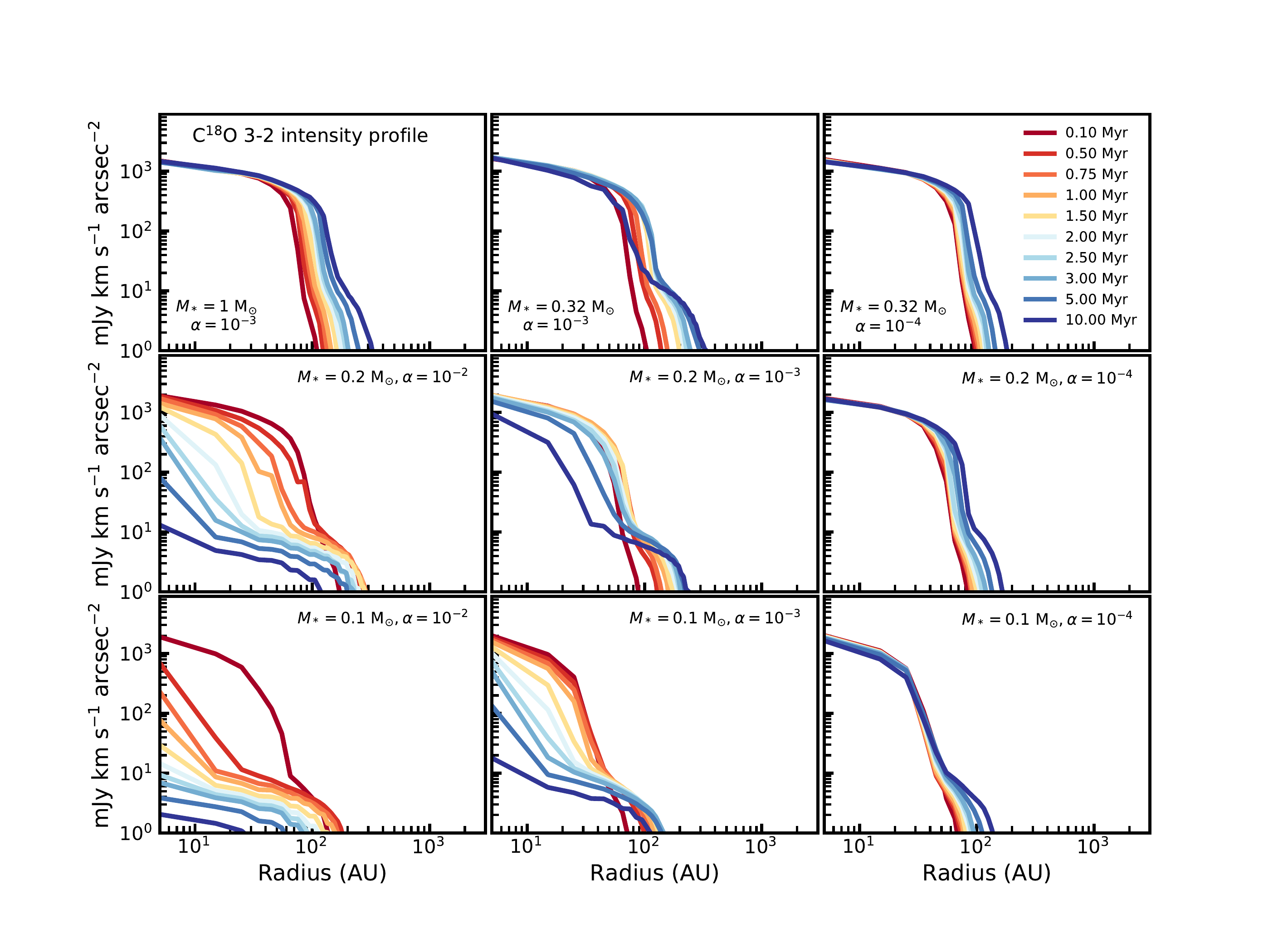}
    \caption{\label{fig: C18O profiles}Time evolution of the \xco\ $J=3\,-\,2$ intensity profiles of our models, where colors denote different time steps. Each panel shows a different \mstar\ and \alp.}
\end{figure*}

\section{Model \xco\ and \cyo\ $J=3\,-\,2$ integrated fluxes}
\label{app: all 3-2 fluxes}

\begin{figure*}
    \centering
    \begin{subfigure}{0.48\textwidth}
    \includegraphics[width=\textwidth]{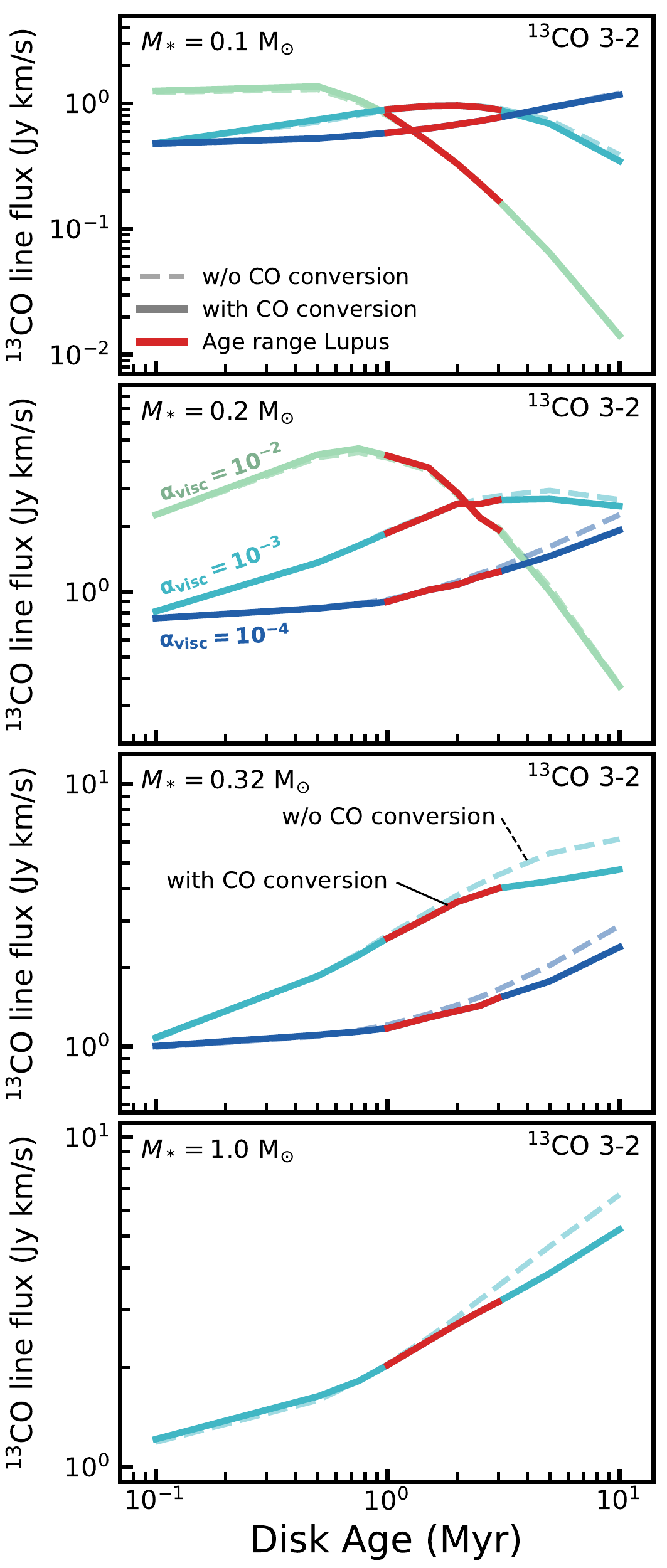}
    \end{subfigure}
    \begin{subfigure}{0.48\textwidth}
    \includegraphics[width=\textwidth]{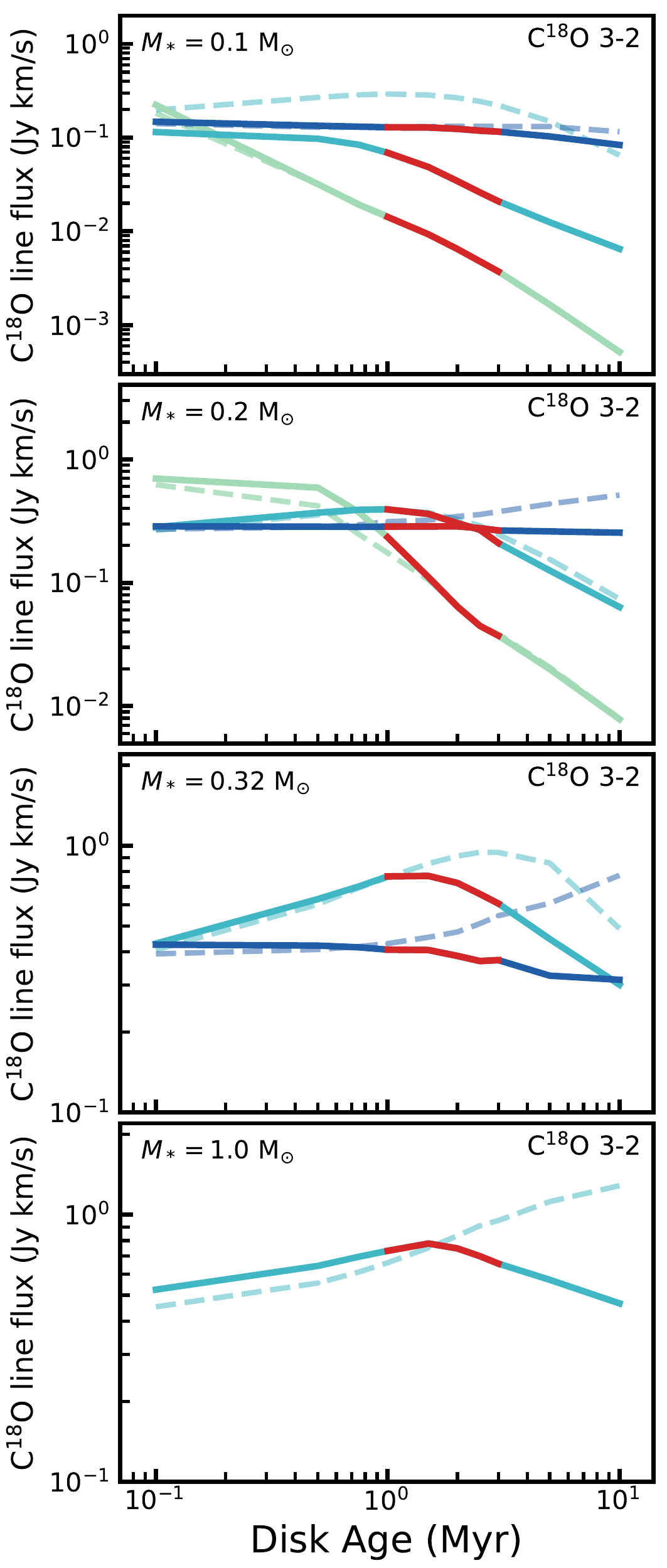}
    \end{subfigure}
    \caption{\label{fig: all xCyO 3-2 fluxes} As Figure \ref{fig: model xCyO fluxes - with gsc}, but showing the \xco\ (left) and \cyo\ (right) $J=3-2$ fluxes for all models. Colors indicate different \alp. Models with and without CO conversion through grain-surface chemistry are shown with solid and dashed lines, respectively. The age range for disks in Lupus is highlighted in red. Few models show a significant decrease in \xco\ line fluxes, except for models with $\alp = 10^{-2}$ that have low disk masses. The \cyo\ line fluxes do show a decrease with age, especially at later disk ages.} 
\end{figure*}

\section{Model \xco\ and \cyo\ $J=2\,-\,1$ fluxes}
\label{app: all 2-1 fluxes}

\begin{figure*}
    \centering
    \begin{subfigure}{0.48\textwidth}
    \includegraphics[width=\textwidth]{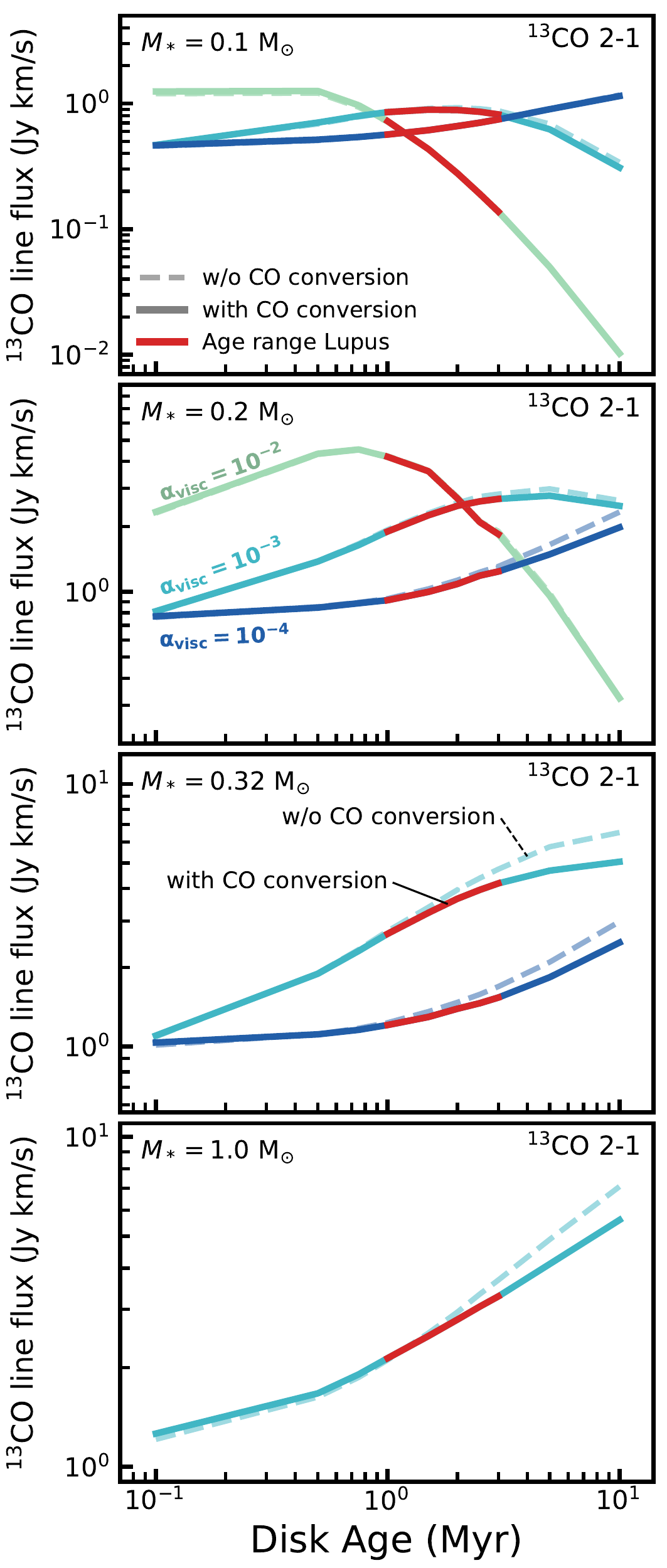}
    \end{subfigure}
    \begin{subfigure}{0.48\textwidth}
    \includegraphics[width=\textwidth]{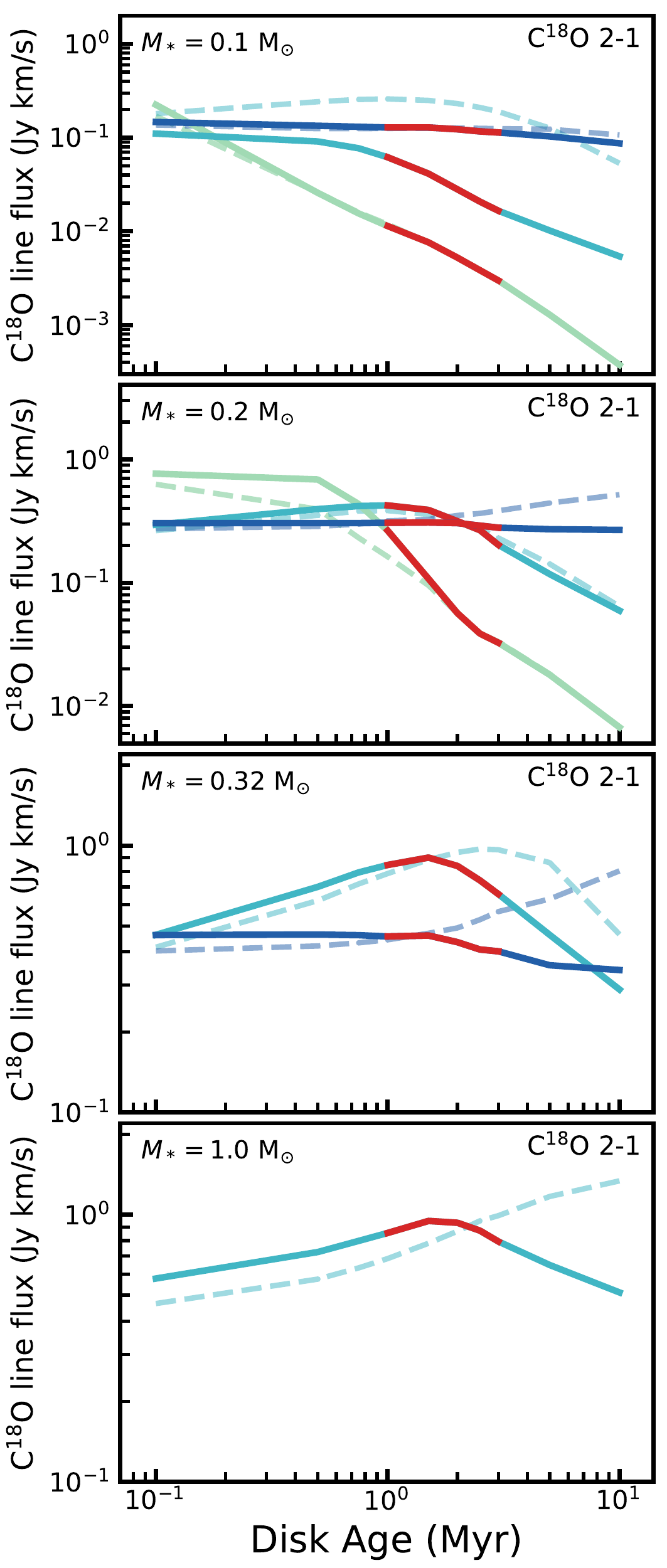}
    \end{subfigure}
    \caption{\label{fig: all xCyO 2-1 fluxes} As Figure \ref{fig: all xCyO 3-2 fluxes}, but showing the \xco\ (left) and \cyo\ (right) $J=2-1$ fluxes for all models. We note that the range of the y-axes are the same as for Figure \ref{fig: all xCyO 3-2 fluxes}. }
\end{figure*}

\section{Comparing maximum CO conversion models to observed CO isotopolog line fluxes}
\label{app: extended version zeta estimation}
\begin{figure*}[b!]
    \centering
    \begin{subfigure}{0.98\textwidth}
    \includegraphics[width=\textwidth]{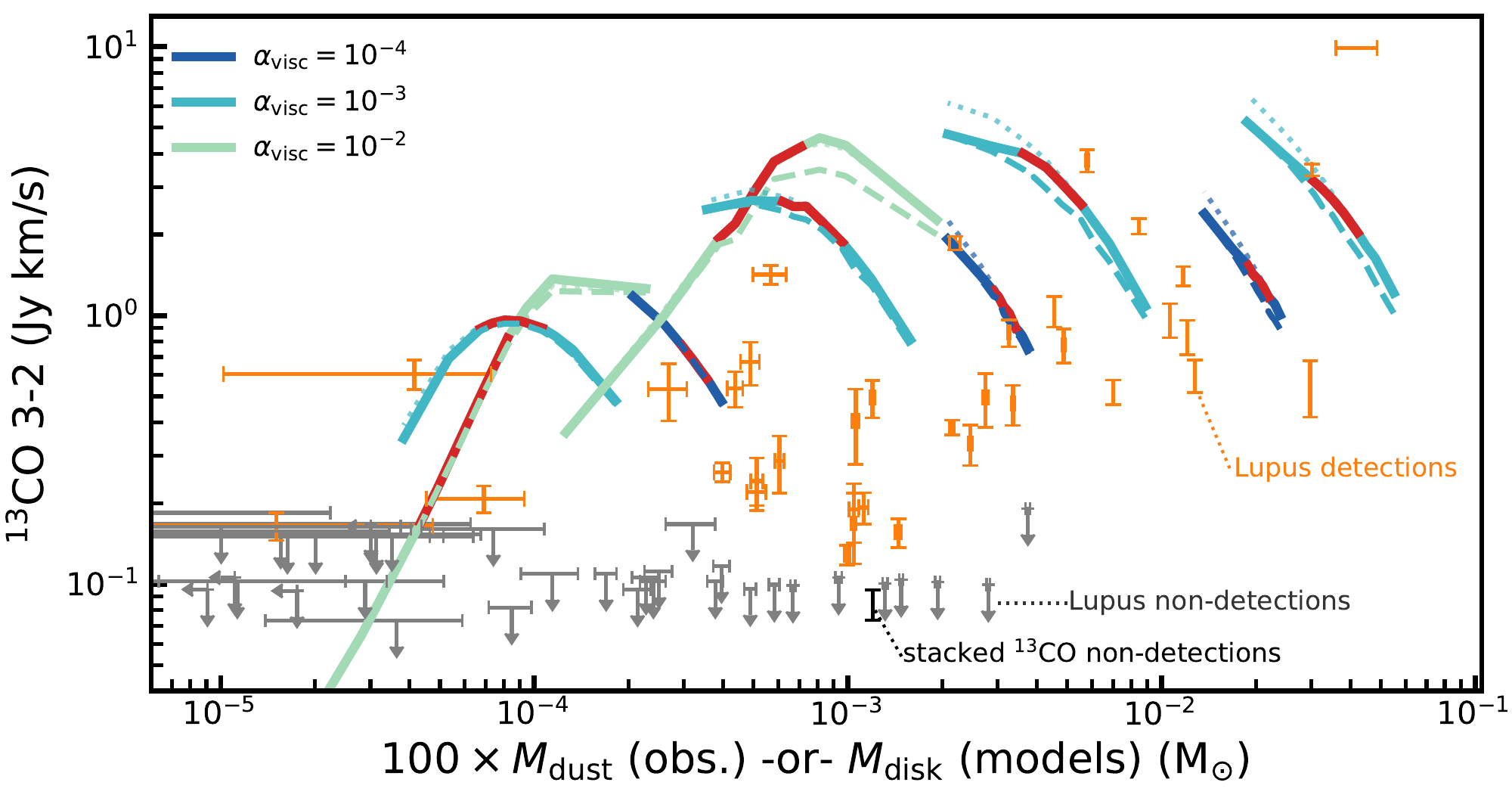}
    \end{subfigure}
    \begin{subfigure}{0.98\textwidth}
    \includegraphics[width=\textwidth]{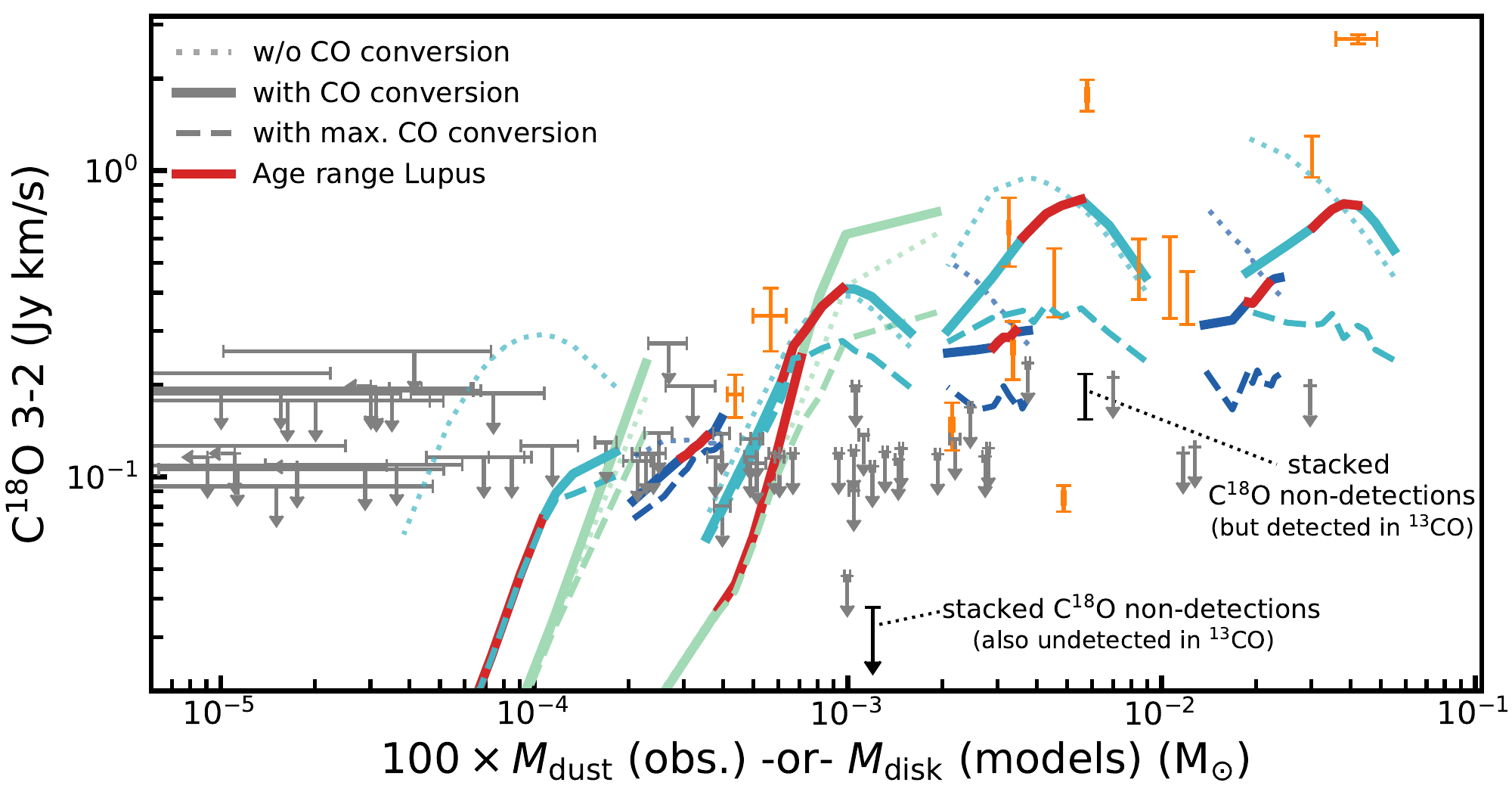}
    \end{subfigure}
    \caption{\label{fig: xCyO fluxes (iso + gsc-INF) vs mdust}
    As Figure \ref{fig: model xCyO fluxes - with gsc}, but now showing three sets of models. The dotted lines show the $^{13}$CO (top) and C$^{18}$O (bottom) $J=3\,-\,2$ line fluxes for models with standard DALI chemistry, that is, without CO conversion. The solid lines show line fluxes for the models where CO has been chemically converted into other species, as described in Section \ref{sec: grain surface chemistry}. The dashed lines show line fluxes for models with maximal CO conversion, where we have removed all CO in the region where CO conversion through grain-surface chemistry occurs (see also Section \ref{fig: zeta estimation zoom in}). Observations in Lupus are shown in orange if detected and gray if an upper limit \citep{ansdell2016,Yen2018}. Stacked non-detections are shown in black (see Section \ref{sec: comparing to observations}).
    }
\end{figure*}

\end{appendix}
\end{document}